\documentclass[aps,floatfix,pra,nofootinbib]{revtex4}
\usepackage{graphicx,bm,color}
\usepackage{array}
\usepackage{amsmath,amsfonts,amssymb,stackrel}
\usepackage{pdflscape}
\usepackage{hyperref,natbib}
\usepackage{url}
\hypersetup{colorlinks=true}
\hypersetup{urlcolor=blue}
\hypersetup{citecolor=black}
\hypersetup{menucolor=black}
\hypersetup{linkcolor=black}
\usepackage{cleveref}
\usepackage{longtable}
\usepackage{float}
\usepackage{pdfpages}
\usepackage[T1]{fontenc}
\usepackage[utf8]{inputenc}
\usepackage{ulem}
\usepackage{multirow}

\newcommand{\be}{\begin{equation}}
\newcommand{\ee}{\end{equation}}
\newcommand{\bmul}{\begin{multline}}
\newcommand{\emul}{\end{multline}}
\newcommand{\bea}{\begin{eqnarray}}
\newcommand{\eea}{\end{eqnarray}}
\newcommand{\rr}{\mathbf{r}}
\newcommand{\kk}{\mathbf{k}}
\newcommand{\qq}{\mathbf{q}}
\newcommand{\kq}[1]{\hat{#1}_\kk^\qq}
\newcommand{\kqb}[1]{\hat{\bar{#1}}_\kk^\qq}

\newcommand{\zero}{\mathbf{0}}

\newcommand{\ii}{{i}}
\newcommand{\dd}{{d}}
\newcommand{\eee}{{e}}

\newcommand{\meanv}[1]{\langle #1 \rangle}

\newcommand{\meanvlr}[1]{\left\langle #1 \right\rangle}

\newcommand{\bb}[1]{\left( #1 \right)}

\newcommand{\bbcro}[1]{\left[ #1 \right]}
\newcommand{\bbcror}[1]{\left. #1 \right]}
\newcommand{\bbcrol}[1]{\left[ #1 \right.}

\newcommand{\deriv}[1]{\ii \hbar \frac{\dd #1}{\dd t}}

\newcommand{\res}[1]{{ #1}}

\DeclareMathOperator\ch{ch}

\begin{document}
\title{Linear response of a superfluid Fermi gas inside its pair-breaking continuum}

\author{H. Kurkjian}
\email{hadrien.kurkjian@gmail.com}
\author{J. Tempere}
\author{S. N. Klimin}

\affiliation{TQC, Universiteit Antwerpen, Universiteitsplein 1, B-2610 Antwerpen, Belgi\"e}

\begin{abstract}
We study the signatures of the collective modes of a superfluid Fermi gas in its linear 
response functions for the order-parameter and density fluctuations in the Random Phase Approximation (RPA).
We show that a resonance associated to the Popov-Andrianov (or sometimes ``Higgs'') mode
is visible inside the pair-breaking continuum  at all values of the wavevector $q$, 
not only in the (order-parameter) modulus-modulus response function but also 
in the modulus-density and density-density responses. At nonzero temperature,
the resonance survives in the presence of thermally broken pairs 
even until the vicinity of the critical temperature $T_c$, and coexists with both the
Anderson-Bogoliubov modes at temperatures comparable to the gap $\Delta$
and with the low-velocity phononic mode predicted by RPA near $T_c$.
The existence of a Popov-Andrianov-``Higgs'' resonance 
is thus a robust, generic feature of the high-energy phenomenology
of pair-condensed Fermi gases, and should be accessible
to state-of-the-art cold atom experiments.
\end{abstract}
\maketitle

\section{Introduction}

A primary way to probe the collective mode spectrum of a many-body system
is by measuring the response functions of its macroscopic observables such as
its density, or, in the case of a condensed system, its order parameter.
These response functions can be measured by driving the system at a given
wavenumber $q$ and varying the drive frequency $\omega$.
In the theoretical case where the collective mode is undamped, one expects
a infinitely narrow resonance (a Dirac peak) when $\omega$
coincides with the collective mode frequency $\omega_\qq$.
However, in most systems, collective modes are coupled to one
or several continua of excitations, for example by intrinsic
couplings to other elementary excitations. 
The system response in this case is less abrupt:
the response functions are nonzero at all frequencies $\omega$ belonging to
the continuum and the Dirac peak of the collective mode is replaced, in the favorable cases,
by a broadened resonance. Theoretically, this damped resonance can be related
to the existence of a pole in the analytic continuation of the response functions 
through their branch cuts associated to the continua \cite{FetterWalecka,Nozieres1963,CohenChapIIIC}.
\res{Eventually, if the coupling to the continuum is very strong, the resonance may entirely disappear,
such that only a slowly varying response remains visible inside the continuum.}

Superfluid Fermi gases, which one can form by cooling down fermionic atoms
prepared in two internal states $\uparrow/\downarrow$ 
\cite{Jin2003,Ketterle2003,Grimm2003,Ketterle2005,Thomas2007,Ketterle2008,Salomon2010,Zwierlein2012,Stringari2013,Vale2017}, 
offer a striking example of this fundamental many-body phenomenon.
This system of condensed pairs of $\uparrow/\downarrow$ fermions is described by 3 collective fields:
the total density $\rho$ of particles and the phase and modulus of the order-parameter $\Delta$.
In the general case, the fluctuations of those 3 fields are coupled and the collective modes have components
on all of them. The system has also fermionic quasiparticles describing the breaking of pairs into unpaired fermions
\cite{BCS1957,Haussmann1993,Zwerger2009,sennepol}, and two fermionic continua of quasiparticle biexcitations: a gapped quasiparticle-quasiparticle
continuum and a gapless quasiparticle-quasihole continuum (to which the collective modes are coupled only at nonzero
temperature). Since the coupling to these continua is not small in general, the collective mode spectrum can be obtained
only after nonperturbative analytic continuations \cite{Schon1975,Popov1976,higgs,artlongsk}.
Performing an analytic continuation to study collective modes coupled to a continuum
is a powerful heuristic tool: it is indispensable to interpret the shape of the response
functions in terms of collective phenomena and to define precisely
the spectrum of the collective branches. However,
the poles found in the analytic continuation are not directly observable and one should
always relate them to resonances which experiments can measure in the response functions.

Meanwhile, the experimental study of the collective modes of a superfluid Fermi gas
is a very active field of research \cite{Thomas2007,Vale2017,Zwierlein2019},
with a special focus on the high-energy collective modes \cite{Koehl2018} (at $\omega$ larger than the quasiparticle-quasiparticle
continuum threshold at $2\Delta$) where a branch with quadratic dispersion \cite{Popov1976,higgs} is expected,
reminiscent of the Higgs modes in high-energy physics \cite{Varma2015},
superconductors \cite{Klein1980,Shimano2013,Sacuto2014,Benfatto2015,Measson2018,Measson2019},
superfluid fermionic Helium \cite{Volovik2014} and nuclear matter \cite{Matera2010,Matera2014}.
This motivates us to discuss the observability, 
in the order-parameter and density response functions of the gas,
of the collective modes predicted by Ref.~\cite{Popov1976,higgs,higgslong,artlongsk}
based on the analytic structure of the functions continuated to imaginary frequencies.
\res{There are two major obstacles \cite{Benfatto2015,Nikuni2013,Varma2015} to the observation of the Popov-Andrianov-``Higgs''
resonance in a conventional fermionic
condensate. $(i)$ So far the resonance has been clearly identified
only in the modulus-modulus response function, whereas experiments (both in superconductors \cite{Varma2015}
and ultracold Fermi gases \cite{Vale2017,Zwierlein2019}) usually excite
or measure the density of the fermions. $(ii)$ In a conventional 
fermionic condensate, where the resonance energy is above $2\Delta$ and
the resonance broadened by its coupling to the pair-breaking
continuum, it is generally not known whether a quality factor and spectral weight large
enough to allow for an observation can be reached. Most studies 
then look for situations where the damping by the continuum is absent,
as in Charged-Density-Wave superconductors 
\cite{Klein1980,Sacuto2014,Benfatto2015,Measson2018,Measson2019}, inhomogeneous systems \cite{Bruun2014} or superfluids 
in unconventional lattice geometries \cite{Nikuni2013}. Here, we show that the resonance
is observable in the density-density and density-modulus response functions at strong coupling.
In those density responses, the spectral weights of the resonance tends to zero with the wavevector $q$ 
while the quality factor decreases when $q$ increases. Nevertheless we could identify an intermediary regime
($q\approx\sqrt{2m\mu}$ at unitarity) where the resonance, and the characteristic quadratic dependence
on $q$ of its peak frequency, should be resolvable from the continuum background in an ultracold Fermi gas.}

We study the response functions in Anderson's Random Phase Approximation (RPA) \cite{Anderson1958}.
We use the formulation of Ref.~\cite{artmicro} in terms of bilinear quasiparticle operators
that we generalize to nonzero temperature and to the presence of external drive fields.
The RPA captures the coupling of the collective modes to the two fermionic
continua (and the corresponding broadening of the resonances in the response functions)
but neglects other couplings, in particular to the continua of two \cite{Beliaev1958}, three
\cite{Khalatnikov1949,annalen} or more collective excitations.
We show that in this approximation, the density fluctuations are sensitive to the fluctuations of $\Delta$, so that
both modulus and phase collective modes are visible in the density response, but that 
the converse is not true. We give explicit expressions of each element of the response function matrix
\cite{FetterWalecka,Takada1988,Mottelson2001,Castin2001}, and show that they agree
with path-integral based treatments \cite{He2016,artlongsk}.

As the spectrum and response-function signatures of the low-energy collective modes is known in the RPA 
at zero \cite{Anderson1958,Strinati1998,CKS2006}, nonzero temperature
\cite{Orbach1981,artmicro} and near the critical temperature $T_c$ \cite{Popov1976,Takada1997,artlongsk},
we concentrate here on the high-energy ($\omega>2\Delta$) modes.
\res{At zero temperature, we show that the resonance of the Popov-Andrianov-``Higgs''
mode is visible not only in the modulus-modulus response \cite{higgs} 
but also as a global extremum (in the region $\omega>2\Delta$)
in the modulus-phase and modulus-density responses, and as a local extremum 
in the density-density response at strong coupling.}
As suggested by the analytic structure found in Ref.~\cite{higgslong}, we show that the branch remains observable
at large $q$ (in particular at $q\approx\sqrt{2m\mu}$ in the weak-coupling limit $\Delta\ll\mu$)
with a quality factor below, but not much below unity. 

At nonzero temperature, where the RPA captures the thermal population 
of the fermionic quasiparticle branches (and only of those branches) \res{and describes
the collective modes in the collisionless approximation}, 
we show that the Popov-Andrianov resonance is
not destroyed by the presence of thermally excited fermionic quasiparticles. 
\res{On the contrary, the increase of temperature (which reduces $\Delta$) favours
the observability of the resonance in the density response functions by increasing
the resonance spectral weight. The shape of the resonance is weakly affected by temperature,
and for the order-parameter responses this shape is actually the same
as at zero temperature for a slightly different interaction strength.
Close to the critical temperature $T_c$,  we show that collective mode
in the pair-breaking continuum branch is not hidden by the low-velocity phononic
branch \cite{artlongsk} as long as $\hbar^2 q^2/m\lesssim \Delta^2 /\mu$. This is in contrast with 
the Anderson-Bogoliubov branch, which disappears near $T_c$ according to the RPA.}

Altogether our findings confirm the observability of the Popov-Andrianov-``Higgs'' branch,
which appears, after our study, as the strongest feature of the high-energy phenomenology 
of pair-condensed Fermi gases. It is observable in wide ranges of values of the interaction strength, 
exciting wavevector and temperature, and it is only weakly affected by the singularities
caused in the response functions by the structure changes of the fermionic continuum.
We are then optimistic about its observability, especially if the experiments 
can access one of the modulus response functions (modulus-modulus,
modulus-phase or modulus-density).
The modulus of the order-parameter can be excited by a Feshbach modulation of the scattering length \cite{Gurarie2009,higgs}, 
after which the modulus-density response can be measured by absorption images as in Ref.~\cite{Vale2017,Zwierlein2019}.
Alternatively the density can be excited by a Bragg pulse \cite{Vale2017} or by shaking the confinement walls \cite{Zwierlein2019}, and the order-parameter modulus
measured after a bosonization of the Cooper pairs. \res{In the density-density response, 
it would be interesting to see if the peak observed in \cite{Vale2017} above
$2\Delta$ has the characteristic behavior of the Popov-Andrianov-``Higgs’’ mode,
that is, a quadratic dependence on $q$ for both the peak frequency and its width.}

\section{BCS theory at nonzero temperature}

To derive the matrix of linear response functions, we use the formalism of Ref.~\cite{artmicro},
itself based on the RPA approach of Anderson \cite{Anderson1958}, and we generalize it
to the presence of pairing and density exciting fields.
We start by briefly recalling the formalism of BCS theory at nonzero temperature.
In real and momentum space, the Hamiltonian of an isolated gas of fermions in two
internal states $\sigma=\uparrow/\downarrow$ with $s$-wave contact interactions is given by
\bea
\hat{H} &=& l^3 \sum_{\rr,\sigma=\uparrow/\downarrow} \hat{\psi}_\sigma^\dagger(\rr) \left( - \frac{1}{2m}  \Delta_{\rr} -\mu\right)
 \hat{\psi}_\sigma(\rr) + g_0 l^3 \sum_{{\rr}} \hat{\psi}_\uparrow^\dagger({\rr}) \hat{\psi}_\downarrow^\dagger(\rr) \hat{\psi}_\downarrow(\rr){\hat{\psi}_\uparrow}(\rr) \\
 &=&   \sum_{\kk\in \mathcal{D},\sigma=\uparrow/\downarrow} \left(\frac{ k^2}{2m} - \mu \right) \hat{a}_{\kk \sigma}^\dagger
 \hat{a}_{\kk \sigma} + \frac{g_0}{V} \sum_{\kk,\kk',\qq\in \mathcal{D}} \hat{a}_{\kk' \uparrow}^\dagger \hat{a}_{-\kk' -\qq \downarrow}^\dagger \hat{a}_{-\kk-\qq \downarrow} \hat{a}_{\kk \uparrow}.
 \label{H}
\eea
We use from now on the convention $\hbar=k_{\rm B}=1$.
To introduce a momentum cutoff in a natural way, we discretize space into a cubic lattice of step $l$ 
and impose periodic boundary conditions (in a volume $V=L^3$), which restrict the values of the wavevectors to 
$\mathcal{D}=\frac{2\pi}{L}\mathbb{Z}^{3}\cap[-\pi/{l},\pi/{l}[^3$.
The bare coupling constant $g_0$ is renormalized to reproduce the correct
$s$-wave scattering length of the two-body problem:
\be
\frac{1}{g_0}=\frac {m}{4 \pi \hbar^2 a}-\int_{[-\pi/l,\pi/l[^3} \frac{\dd^3 k}{(2\pi)^3} \frac{m}{{ k^2}}.
\label{g0_a}
\ee
At the end of the calculation we take the lattice spacing $l$ to $0$, and thus $g_0$ tends to $0$ to compensate the divergence 
of the integral on the right-hand-side of \eqref{g0_a}.

BCS theory describes the equilibrium state at temperature $T$ by the Gaussian state:
\be
\hat{\rho}_{\rm BCS}(T)=\frac{\exp\bb{-\hat H_{\rm BCS}/ T}}{\mathcal{Z}}
\label{rhoT}
\ee
where $\mathcal{Z}$ is the partition function and the BCS Hamiltonian $H_{\rm BCS}$ is obtained by treating the interactions in the mean-field approximation, \textit{i.e.}
by replacing the quartic interaction
term $g_0 \hat{\psi}_\uparrow^\dagger({\rr}) \hat{\psi}_\downarrow^\dagger(\rr) \hat{\psi}_\downarrow(\rr) \hat{\psi}_\uparrow(\rr)$ in \eqref{H}
by a quadratic one $ \Delta (\hat{\psi}_\uparrow^\dagger({\rr}) \hat{\psi}_\downarrow^\dagger(\rr) +\textrm{cc})$, 
through the introduction of the self-consistent pairing-field
\be
\Delta=g_0\meanvlr{\hat{\psi}_\uparrow^\dagger({\rr}) \hat{\psi}_\downarrow^\dagger(\rr)}_T.
\ee
Here $\meanv{\ldots}_T$ denotes the average in the thermal state $\hat{\rho}_{\rm BCS}(T)$. This quadratic Hamiltonian
can be diagonalized easily into a Hamiltonian describing fermionic elementary excitations on top of a ground state energy $E_0$:
\be
\hat{H}_{\rm BCS}=E_0+\sum_{\kk\in\mathcal{D}} \epsilon_\kk \hat\gamma_{\kk\sigma}^\dagger \hat\gamma_{\kk\sigma}.
\ee
Here the eigenenergy of a fermionic excitation is
\be
\epsilon_\kk=\sqrt{\xi_\kk^2+\Delta^2} \quad \mbox{with} \quad \xi_\kk=\frac{k^2}{2m}-\mu.
\ee
The fermionic quasiparticle operators $\hat\gamma_{\kk,\sigma}$
are obtain after a Bogoliubov rotation of the particle operators 
$\hat a_{\kk,\sigma}$ as in the zero temperature case:
\bea
\hat{{\gamma}}_{\kk\uparrow}     &=& U_\kk \hat{a}_{\kk\uparrow} + V_\kk \hat{a}_{-\kk\downarrow}^\dagger \label{gamup} \\
\hat{{\gamma}}_{-\kk\downarrow} &=& -V_\kk \hat{a}_{\kk\uparrow}^\dagger + U_\kk \hat{a}_{-\kk\downarrow}  \label{gamdown}
\eea
with the Bogoliubov coefficients $U_\kk$ and $V_\kk$:
\be
U_\kk=\sqrt{1+\frac{\xi_\kk}{\epsilon_\kk}} \quad \mbox{and} \quad V_\kk=\sqrt{1-\frac{\xi_\kk}{\epsilon_\kk}}.
\ee
The difference with the zero temperature case lies in the average values of the bilinear operators:
\bea
\meanv{\hat a^\dagger_{\kk,\sigma}\hat a_{\kk,\sigma}} &=& \bb{U_\kk^2-V_\kk^2} f_\kk+V_\kk^2 \\
\meanv{\hat a_{-\kk,\downarrow}\hat a_{\kk,\uparrow}} &=& -\bb{1-2f_\kk}U_\kk V_\kk,
\eea
which now depend on the Fermi-Dirac occupation number
\be
f_\kk=\meanv{\hat \gamma^\dagger_{\kk,\sigma}\hat \gamma_{\kk,\sigma}}_{T}=\frac{1}{1+\exp{\epsilon_\kk/T}}.
\ee
This thermal population of the quasiparticle modes also affect the gap equation:
\be
\Delta=-\frac{g_0}{L^3}\sum_{\kk\in\mathcal{D}} U_\kk V_\kk (1-2f_\kk).
\label{Delta}
\ee
Thus, at nonzero temperature, BCS theory captures the effects due to the thermally excited fermionic quasiparticles (the broken
pairs); it completely neglects that there are also thermally excited collective modes (some of which are
gapless) which is a serious limitation, particularly at strong coupling.

\section{RPA equations of motion in presence of drive fields}

To study the linear response of the gas, we introduce, on top of the Hamiltonian \eqref{H} of the isolated gas, 
a quadratic Hamiltonian describing the experimental driving of the system:
\be
\hat{H}_{\rm drive} = l^3\sum_\rr \bb{  u_\uparrow (\rr,t) \hat\psi_\uparrow^\dagger(\rr)  \hat\psi_\uparrow(\rr) +  u_\downarrow (\rr,t) \hat\psi_\downarrow^\dagger(\rr) \hat\psi_\downarrow(\rr) + \bbcro{  \hat\psi^\dagger_\uparrow(\rr)\hat\psi^\dagger_\downarrow(\rr) \phi(\rr,t)+\mbox{c.c.}}}.
\ee
Here the fields $u_\sigma(\rr)$, coupled to the density of spin $\sigma$ fermions, describe for instance a Bragg excitation
of the gas \cite{Vale2017}. The complex field $\phi(\rr)$ coupled to the quantum pairing field $\hat\psi^\dagger_\uparrow(\rr)\hat\psi^\dagger_\downarrow(\rr)$ can be imposed for instance by a Feshbach-modulation of the interaction strength \cite{Gurarie2009}. An excitation
coupled to the phase of $\hat\psi^\dagger_\uparrow(\rr)\hat\psi^\dagger_\downarrow(\rr)$ can be achieved using a time- and space-dependent 
Josephson junction as proposed in \cite{artlongsk}. This drive Hamiltonian decomposes into a sum of Fourier components of the momentum $\qq$ 
transferred to system:
\be
\hat{H}_{\rm drive} = \sum_\qq \hat{H}_{\rm drive}(\qq) \qquad \mbox{with} \qquad \hat{H}_{\rm drive}(\qq)= \sum_\kk \bb{u_\uparrow(-\qq)  \hat n_\kk^\qq+u_\downarrow(-\qq) \hat{\bar n}_\kk^\qq} + \phi(-\qq)  \sum_\kk  \hat{\bar{d}}_\kk^\qq + \bar\phi(-\qq) \sum_\kk  \hat{{d}}_\kk^\qq.
\ee
We use here Anderson's \cite{Anderson1958} notations for the bilinear fermion operators\footnote{See also chapter V in \cite{TheseHK}. Note the
symmetrization of the notation of the fermion momenta $\kk\pm\qq/2$.},
\begin{align}
 \hat{n}_{\kk}^\qq &= \hat{a}_{\kk+\qq/2 \uparrow}^\dagger \hat{a}_{\kk-\qq/2 \uparrow}   &\hat{\bar{n}}_{\kk}^{\qq} &= \hat{a}_{-\kk+\qq/2 \downarrow}^\dagger \hat{a}_{-\kk-\qq/2 \downarrow} \notag \\
 \hat{d}_\kk^\qq &= \hat{a}_{-\kk-\qq/2 \downarrow} \hat{a}_{\kk-\qq/2 \uparrow}   &\hat{\bar{d}}_\kk^{\qq} &= \hat{a}_{\kk+\qq/2 \uparrow}^\dagger \hat{a}_{-\kk+\qq/2 \downarrow}^\dagger
\label{bilinop}
\end{align}
and the Fourier transforms of the drive fields:
\bea
u_\sigma(\qq)&=&\frac{l^3}{V} \sum_\rr \eee^{\ii\qq\cdot\rr} u_\sigma(\rr) \\
\phi(\qq)&=&\frac{l^3}{V} \sum_\rr \eee^{\ii\qq\cdot\rr} \phi(\rr) \quad \mbox{and} \quad \bar\phi(\qq) =\phi^*(-\qq).
\eea

In the framework of linear response theory, we seek the response of the system to first order in the fields $\phi$ and $u_\sigma$. We thus neglect the quantum fluctuations in the terms of the equations of motion deriving from $\hat{H}_{\rm drive}$: 
\be
[\hat a \hat b,\hat{H}_{\rm drive}]\simeq\meanvlr{[\hat a \hat b,\hat{H}_{\rm drive}]}_T,
\ee 
\res{where the average value  $\meanv{{[\hat a \hat b,\hat{H}_{\rm drive}]}}_T$ is taken in the BCS equilibrium state at zero fields.}
The rest of the derivation is similar to what is explained in Refs.~\cite{TheseHK,artmicro}: one writes the Heisenberg
equations of motion for the bilinear fermionic operators \eqref{bilinop} and linearizes them using incomplete
Wick contractions (\textit{i.e} the replacement $\hat{a} \hat{b} \hat{c} \hat{d} \rightarrow \hat{a}\hat{b} \langle\hat{c}\hat{d} \rangle_T  + \langle\hat{a}\hat{b} \rangle_T \hat{c}\hat{d} -\hat{a}\hat{c} \langle\hat{b}\hat{d}\rangle_T -  \langle\hat{a}\hat{c}\rangle_T \hat{b}\hat{d} + \ldots$). The resulting equations of motion
of the bilinear particle operators are given in Appendix \ref{app:eom}. We give here the equations of motion in their simplest form, which is in the quasiparticle basis. At the level of the bilinear operators, the Bogoliubov rotation \eqref{gamup}--\eqref{gamdown} becomes:
\be
\begin{pmatrix}
\kq{y} \\
\kq{h} \\
\kq{s} \\
\kq{m}
\end{pmatrix}
\equiv
\begin{pmatrix}
 \hat{\gamma}_{-\kk_+,\downarrow} \hat{\gamma}_{\kk_-,\uparrow}    -     \hat{\gamma}_{\kk_+,\uparrow}^\dagger \hat{\gamma}_{-\kk_-,\downarrow}^\dagger \\
\hat{\gamma}_{\kk_+ \uparrow}^\dagger \hat{\gamma}_{\kk_- \uparrow} - \hat{\gamma}_{-\kk_+ \downarrow}^\dagger \hat{\gamma}_{-\kk_- \downarrow} \\
 \hat{\gamma}_{-\kk_+,\downarrow} \hat{\gamma}_{\kk_-,\uparrow}    +     \hat{\gamma}_{\kk_+,\uparrow}^\dagger \hat{\gamma}_{-\kk_-,\downarrow}^\dagger  \\
\hat{\gamma}_{\kk_+ \uparrow}^\dagger \hat{\gamma}_{\kk_- \uparrow} + \hat{\gamma}_{-\kk_+ \downarrow}^\dagger \hat{\gamma}_{-\kk_- \downarrow} \\
\end{pmatrix}
=
\begin{pmatrix}
W_{\kk\qq}^+ & w_{\kk\qq}^- & 0&0  \\
-w_{\kk\qq}^-  &   W_{\kk\qq}^+ & 0&0  \\
0&0&  W_{\kk\qq}^-  &- w_{\kk\qq}^+ \\
0&0& w_{\kk\qq}^+   &  W_{\kk\qq}^-   \\
\end{pmatrix}
\begin{pmatrix}
\kq{d} - \kq{\bar{d}}\\
\kq{n} - \kq{\bar{n}}\\
\delta (\kq{d} + \kq{\bar{d}})\\
\delta (\kq{n} + \kq{\bar{n}})\\
\end{pmatrix},
\label{passage_qpart}
\ee
where we have used $\kk_\pm=\kk\pm\qq/2$, $W_{\kk\qq}^\pm=U_{\kk_+}U_{\kk_-}\pm V_{\kk_+}V_{\kk_-}$ and $w_{\kk\qq}^\pm=U_{\kk_+}V_{\kk_-}\pm V_{\kk_+}U_{\kk_-}$. Performing this change of basis on the equations of motion, we get:
\begin{align}
\deriv{\kq{y}}&= \epsilon_{\kk\qq}^+ \kq{s} +  (1-f_{\kk_+}-f_{\kk_-}) \bbcro{W_{\kk\qq}^- \bb{\delta\hat{\Delta}^\qq+\delta\hat{\bar{\Delta}}^\qq + \phi_+(\qq)} -  w_{\kk\qq}^+ \bb{g_0\bbcro{\delta\hat{n}_\uparrow^\qq+\delta\hat{{n}}_\downarrow^\qq}+u_+(\qq)}} \label{systRPA1} \\
\deriv{\kq{s}}&= \epsilon_{\kk\qq}^+ \kq{y} + (1-f_{\kk_+}-f_{\kk_-}) \bbcro{W_{\kk\qq}^+  \bb{\hat{\Delta}^\qq-\hat{\bar{\Delta}}^\qq+\phi_-(\qq)} -  w_{\kk\qq}^- \bb{g_0\bbcro{\hat{n}_\uparrow^\qq-\hat{{n}}_\downarrow^\qq}-u_-(\qq)} } \label{systRPA2} \\
\deriv{\kq{m}}&=-\epsilon_{\kk\qq}^- \kq{h} - (f_{\kk_+}-f_{\kk_-}) \bbcro{w_{\kk\qq}^- \bb{\hat{\Delta}^\qq-\hat{\bar{\Delta}}^\qq+\phi_-(\qq)} +  W_{\kk\qq}^+ \bb{g_0\bbcro{\hat{n}_\uparrow^\qq-\hat{{n}}_\downarrow^\qq}-u_-(\qq)} }  \label{systRPA3} \\
\deriv{\kq{h}}&=-\epsilon_{\kk\qq}^- \kq{m}+  (f_{\kk_+}-f_{\kk_-})  \bbcro{w_{\kk\qq}^+ \bb{\delta\hat{\Delta}^\qq+\delta\hat{\bar{\Delta}}^\qq + \phi_+(\qq)}+ W_{\kk\qq}^- \bb{g_0\bbcro{\delta\hat{n}_\uparrow^\qq+\delta\hat{{n}}_\downarrow^\qq}+u_+(\qq)}}, \label{systRPA4}
\end{align}
where $\phi_\pm(\qq) = \phi(\qq)\pm\bar\phi(\qq)$, $u_\pm(\qq) = u_\uparrow(\qq)\pm u_\downarrow(\qq)$ and\footnote{This subtraction of the mean-field average value matters only for the $q=0$ system.} $\delta\hat O=\hat O-\meanv{\hat O}_T$. 
At the linear order, the sole effect of the drive fields is thus to shift the collective quantities 
which enter in the equations of motion:
\bea
\hat{\Delta}^\qq                     &=& \frac{g_0}{L^3}\sum_{\kk_1} \hat{{d}}_{\kk_1}^{\qq} \quad \to \quad \hat{\Delta}^\qq+\phi(\qq)  \label{eq:op_col_1} \\
\hat{\bar{\Delta}}^\qq            &=& \frac{g_0}{L^3} \sum_{\kk_1} \hat{\bar{d}}_{\kk_1}^{\qq} \quad \to \quad \hat{\bar{\Delta}}^\qq+\bar\phi(\qq) \label{eq:op_col_2} \\
\hat{n}^{\qq}_\uparrow     &=& \frac{1}{L^3}\sum_{\kk_1} \hat{n}_{\kk_1}^{\qq} \quad \to \quad \hat{n}^{\qq}_\uparrow +u_\downarrow(\qq) /g_0 \label{eq:op_col_3} \\
\hat{n}^{\qq}_\downarrow &=& \frac{1}{L^3} \sum_{\kk_1} \hat{\bar{n}}_{\kk_1}^{\qq} \quad \to \quad\hat{n}^{\qq}_\downarrow+u_\uparrow(\qq)/g_0.  \label{eq:op_col_4}
\eea
Note that one recovers the zero temperature system Eqs.~(14--16) of \cite{artmicro} by setting $f_\kk=0$
(in which case the equations of motion of the $\hat \gamma^\dagger \hat \gamma$ operators become trival).

\section{Linear response to a periodic drive}

\subsection{Matrix of response functions of a driven system}
We now assume that the system is driven at a fixed frequency $\omega$, such that $\phi(\rr,t)=\phi(\rr)\eee^{\ii\omega t}$ and $u_\sigma(\rr,t)=u_\sigma(\rr)\eee^{\ii\omega t}$. We can then replace the time derivatives $\ii\hbar\partial_t$ in Eqs.~(\ref{systRPA1}--\ref{systRPA4}) by $\omega+\ii0^+$. Rederiving with respect to time and resumming the system to form the collective quantities (\ref{eq:op_col_1}--\ref{eq:op_col_4}) yields the $4$-dimensional linear system
\be
\bb{\frac{V}{g_0}\mathbb{I}-\Pi(\omega,\qq)} \begin{pmatrix} 2\ii\Delta\hat\theta^\qq \\  2\delta|\Delta^\qq| \\ g_0\delta\hat\rho^\qq \\ g_0\hat p^\qq\end{pmatrix} = \Pi(\omega,\qq) \begin{pmatrix} \phi_-(\qq) \\ \phi_+(\qq) \\  u_+(\qq) \\ - u_-(\qq)  \end{pmatrix},
\label{4par4}
\ee
where $\mathbb{I}$ is the identity matrix. We have introduced the density and polarisation fluctuations and reparametrized the fluctuations of the order-parameter as:
\bea
\hat{\Delta}^\qq &=& (\Delta+\delta|\Delta^\qq|)\eee^{\ii\hat\theta^\qq} \label{phase}\\
\hat{\bar\Delta}^\qq &=& (\Delta+\delta|\Delta^\qq|)\eee^{-\ii\hat\theta^\qq} \label{module} \\
\hat{\rho}^\qq &=& \hat{n}^{\qq}_\uparrow + \hat{n}^{\qq}_\downarrow \label{densite}\\
\hat{p}^\qq &=& \hat{n}^{\qq}_\uparrow - \hat{n}^{\qq}_\downarrow. \label{polarisation}
\eea
We treat the phase $\hat\theta^\qq$ of the order-parameter as an infinitesimal 
and therefore linearize the exponential in \eqref{phase}--\eqref{module}, 
which is consistent with our symmetry-breaking approach where the expansion 
is done around the mean-field state with a real $\Delta$. The system's linear response 
matrix \cite{FetterWalecka,Takada1988,Mottelson2001,Castin2001}, which relates the fluctuations of the density
and order-parameter to the infinitesimal drive fields, is then
\be
\chi=\frac{\Pi}{\frac{V}{g_0}\mathbb{I}-\Pi}.
\ee
To describe the experimental behavior of a driven system, one usually concentrates
on the imaginary part of $\chi$, which describes the energy absorbed by the system \cite{FetterWalecka}
(whereas the real part describes the energy refracted or reflected by the system).

The matrix $\chi$ is expressed in terms of the $4\times4$ matrix $\Pi$ of the pair correlation functions \cite{Griffin2003},
computed in the BCS thermal state \eqref{rhoT}:
\be
\Pi=
\begin{pmatrix}
\Sigma_{W^+ W^+}^{\epsilon} & \Sigma_{W^- W^+}^{\omega} & -\Sigma_{w^+ W^+}^\omega & 0 \\
\Sigma_{W^+ W^-}^{\omega} & \Sigma_{W^- W^-}^{\epsilon} & - \Sigma_{w^+ W^-}^\epsilon & 0 \\
-\Sigma_{W^+ w^+}^{\omega} & -\Sigma_{W^- w^+}^{\epsilon} & \Sigma_{w^+w^+}^{\epsilon} & 0 \\
0 & 0 & 0 & -\Sigma_{w^-w^-}^{\epsilon}
\end{pmatrix}
-
\begin{pmatrix}
S_{w^-w^-}^{\epsilon} & S_{w^+ w^-}^{\omega} & S_{W^- w^-}^{\omega} & 0\\
S_{w^- w^+}^{\omega} & S_{w^+w^+}^{\epsilon}  &S_{W^-w^+}^{\epsilon} & 0\\
S_{w^- W^-}^{\omega} & S_{w^+ W^-}^{\epsilon} & S_{W^-W^-}^{\epsilon} & 0 \\
0 & 0 & 0 & -S_{W^+ W^+}^{\epsilon}
\end{pmatrix}
\label{Pi}
\ee
where we generalize the notations of Refs.~\cite{TheseHK,higgslong}~:
\begin{align}
&&\Sigma_{ab}^{\epsilon}(\omega,\qq)= \sum_\kk \frac{\epsilon_{\kk\qq}^+ a_{\kk\qq} b_{\kk\qq}(1-f_{\kk_+}-f_{\kk_-})}{\omega^2-(\epsilon_{\kk\qq}^+)^2} 
\qquad &&\Sigma_{ab}^{\omega}(\omega,\qq) = \sum_\kk \frac{\omega a_{\kk\qq} b_{\kk\qq}(1-f_{\kk_+}-f_{\kk_-})}{\omega^2-(\epsilon_{\kk\qq}^+)^2} \label{Sigma} \\
&&S_{ab}^{\epsilon}(\omega,\qq)= \sum_\kk \frac{\epsilon_{\kk\qq}^- a_{\kk\qq} b_{\kk\qq}(f_{\kk_+}-f_{\kk_-})}{\omega^2-(\epsilon_{\kk\qq}^-)^2} 
\qquad  &&S_{ab}^{\omega}(\omega,\qq) = \sum_\kk \frac{\omega a_{\kk\qq} b_{\kk\qq}(f_{\kk_+}-f_{\kk_-})}{\omega^2-(\epsilon_{\kk\qq}^-)^2}. \label{S}
\end{align}
\res{Here $a$ and $b$ are one of the functions $W^+$, $W^-$, $w^+$ or $w^-$ of $\kk$ and $\qq$, and $\epsilon_{\kk\qq}^+$ is short-hand for $\epsilon_{\kk+\qq/2}\pm\epsilon_{\kk-\qq/2}$.}
The first and second matrix in the right-hand side of \eqref{Pi} are the contribution of respectively the quasiparticle-quasiparticle and quasiparticle-quasihole continua to $\Pi$.
In our unpolarized system, the polarization fluctuations $\hat{n}_\uparrow^\qq-\hat{{n}}_\downarrow^\qq$ 
are entirely decoupled from the other collective fields. Note that
the response functions computed here for a driven system also give access, through a Laplace transform \cite{higgslong},
to the time response following a perturbation localized in time.

\res{Remark that, up to some signs, the matrix $\Pi$ has a tensor-product structure when expressed in terms
of the vector $(a_1,a_2,a_3,a_4)=(W^+,W^-,w^+,w^-)$:
\be
\Pi_{ij}=
\begin{cases}
\eta_{ij}\Sigma_{a_i,a_j}^\epsilon-\eta’_{ij}S_{a_{5-i},a_{5-j}}^\epsilon \quad \mbox{if} \quad i=j \quad \mbox{or} \quad i=5-j\\
\eta_{ij}\Sigma_{a_i,a_j}^\omega-\eta’_{ij}S_{a_{5-i},a_{5-j}}^\omega \quad \mbox{else}
\end{cases}
\ee
The signs $\eta_{ij}=\pm1$ and $\eta’_{ij}=\pm1$ should be read on Eq.~\eqref{Pi}.}

\subsection{Eigenenergies of the collective modes}
The response of the system should diverge when 
the drive frequency coincides with the eigenfrequency $\omega_\qq$ of a collective mode;
to find those eigenfrequencies, one should thus search for the poles of $\chi$, in other
words the zero of its denominator\footnote{Note that the bare density-density response function 
$\Pi_{33}$ may have poles in the complex plane, which remain poles of the dressed
response $\chi_{33}$.}:
\be
\mbox{det} \bb{\frac{V}{g_0}\mathbb{I}-\Pi(z_\qq,\qq)} = 0.
\label{modesco}
\ee
When $\Pi$ has a branch cut on the real axis (which occurs for all $\omega \in \mathbb{R}$ at nonzero temperature
and for $|\omega|>\min_{\kk}(\epsilon_{\kk+\qq/2}+\epsilon_{\kk-\qq/2})$ at zero temperature), this equation cannot
have a real solution. Its analytic continuation to the lower-half complex plane may however
have solutions describing damped collective modes. Numerical and analytic methods to continue the matrix $\Pi$
through its branch cuts have been described in \cite{higgs,artlongsk,higgslong}.

\subsection{Explicit expressions of the response functions in the limit $g_0\to0$}
In the limit of zero lattice spacing ($l\to0$), $g_0$ tends to $0$, $\Pi_{11}$ and $\Pi_{22}$ 
are equivalent to $V/g_0$, while $\Pi_{33}$ and $\Pi_{44}$ have a finite limit\footnote{To interpret physically the elements of $\Pi$
the reader can use the correspondance $1,2,3,4$ $\to$ $\theta,|\Delta|,\rho,p$.}.  We thus have the equivalences
\be
\Pi\underset{g_0\to0}{\sim}
\begin{pmatrix}
V/g_0 & \Pi_{12} & \Pi_{13}  & 0 \\
\Pi_{12} & V/g_0 & \Pi_{23} & 0 \\
\Pi_{13}  &\Pi_{23}& \Pi_{33} & 0 \\
0 & 0 & 0 &  \Pi_{44}
\end{pmatrix},
\qquad 
\Pi-\frac{V}{g_0}\mathbb{I}\underset{g_0\to0}{\sim}
\begin{pmatrix}
\Pi_{11} -V/g_0 & \Pi_{12} & \Pi_{13}  & 0 \\
\Pi_{12} & \Pi_{22} -V/g_0 & \Pi_{23} & 0 \\
\Pi_{13}  &\Pi_{23}& -V/g_0 & 0 \\
0 & 0 & 0 & -V/g_0
\end{pmatrix}.
\label{equivalents}
\ee
Note that $\Pi_{11}-V/g_0$ and $\Pi_{22}-V/g_0$ have a finite limit when $g_0\to0$. The determinant of the denominator of $\chi$ is then proportional to the determinant of the $2\times2$ upper left submatrix:
\be
\mbox{det} \bb{\Pi-\frac{V}{g_0}\mathbb{I}}\underset{g_0\to0}{\sim} \bb{\frac{V}{g_0}}^2 \mathcal{D} \qquad \mbox{with} \qquad \mathcal{D}={ \bb{\Pi_{11}-V/g_0} \bb{\Pi_{22}-V/g_0}  - \bb{\Pi_{12}}^2}.\label{D}
\ee
Physically this means that the collective mode spectrum 
is entirely determined by the (modulus and phase) fluctuations of
the order-parameter. However, the density responses may exhibit collective
mode resonances as a result of the density-order parameter couplings.
Using the equivalences \eqref{equivalents} to compute the matrix product in $\chi$ we obtain:
\begin{multline}
\tilde\chi\equiv \begin{pmatrix}1&0&0&0\\0&1&0&0\\0&0&V/g_0&0\\0&0&0&V/g_0\end{pmatrix}
\chi \begin{pmatrix}g_0/V&0&0&0\\0&g_0/V&0&0\\0&0&1&0\\0&0&0&1\end{pmatrix} \underset{g_0\to0}{\sim} 
-\frac{1}{\mathcal D} \\
\times\left(
\begin{tabular}{cccc}
 $\tilde  \Pi_{22}$ & $-\Pi_{12}$  &  $\Pi_{13}\tilde \Pi_{22}-\Pi_{23}\Pi_{12}$ & 0 \\ 
 $-\Pi_{12}$ & $\tilde \Pi_{11}$&  $\Pi_{23}\tilde \Pi_{11}-\Pi_{13}\Pi_{12}$ & 0 \\ \vspace{0.3cm}
\text{{$({\Pi_{13}\tilde \Pi_{22}-\Pi_{12}\Pi_{23}})$}} &  \text{{$({\Pi_{23}\tilde \Pi_{11}-\Pi_{13}\Pi_{12}})$}} & {\small $(\Pi_{13}^2 \tilde \Pi_{22} +  \Pi_{23}^2 \tilde \Pi_{11} -2  \Pi_{13} \Pi_{23} \Pi_{12}- \Pi_{33}\mathcal{D}) $} & 0 \\
0 & 0  &0& $-\Pi_{44} \mathcal{D} $ 
\end{tabular}
\right).
\label{chi}
\end{multline}
We have used the notation $\tilde{\Pi}=\Pi-\frac{V}{g_0}$ and defined the response function matrix $\tilde\chi$ in the basis\footnote{
In this basis Eq.~\eqref{4par4} reads $\begin{pmatrix} 2\ii\Delta\hat\theta^\qq \\  2\delta|\Delta^\qq| \\ V\delta\hat\rho^\qq \\ V\hat p^\qq\end{pmatrix} = \tilde\chi \begin{pmatrix} V\phi_-(\qq)/g_0 \\ V\phi_+(\qq)/g_0 \\  u_+(\qq) \\ - u_-(\qq) \\  \end{pmatrix}$.} where all its coefficients are of order unity when $g_0\to0$. \res{For the density response functions we have in particular:
\bea
\tilde\chi_{13} &=& \frac{\Pi_{23}\Pi_{12}-\Pi_{13}\tilde \Pi_{22}}{  \tilde \Pi_{11} \tilde \Pi_{22}-\Pi_{12}^2} \label{chi13}\\
\tilde\chi_{23} &=& \frac{\Pi_{13}\Pi_{12}-\Pi_{23}\tilde \Pi_{11}}{  \tilde \Pi_{11} \tilde \Pi_{22}-\Pi_{12}^2} \label{chi23}\\
\tilde\chi_{33} &=& \frac{2  \Pi_{13} \Pi_{23} \Pi_{12}-\Pi_{13}^2 \tilde \Pi_{22} -  \Pi_{23}^2 \tilde \Pi_{11} }{  \tilde \Pi_{11} \tilde \Pi_{22}-\Pi_{12}^2} + \Pi_{33}, \label{chi33}
\eea
which coincides with the explicit expressions obtained by Refs.~\cite{He2016,artlongsk} in the path-integral formalism.} At weak coupling ($\Delta/\mu\to0$) and $q=O(\Delta)$, the modulus-phase and modulus-density matrix elements $\Pi_{12}$ and $\Pi_{23}$ vanish such that the collective modes are either pure modulus modes (if their eigenenergy solves $\Pi_{22}(z_\qq)-V/g_0=0$) or pure density-phase modes (if their eigenenergy solves $\Pi_{11}(z_\qq)-V/g_0=0$).

\subsection{Angular points of the response functions}
We conclude this section by remarking that the response functions have the same angular points as
the spectral density
\be
\rho(\omega)=\frac{\Pi(\omega+\ii0^+)-\Pi(\omega-\ii0^+)}{-2\ii\pi}.
\ee
The quasiparticle-quasiparticle part of the spectral density (which originates from the first matrix in \eqref{Pi} with 
$\omega-(\epsilon_{\kk+\qq/2}+\epsilon_{\kk-\qq/2})$ in the denominator) is nonzero
when the ``pair-breaking'' resonance condition is satisfied
\be
\omega=\epsilon_{\kk+\qq/2}+\epsilon_{\kk-\qq/2}.
\ee
Physically, when this resonance condition is met, the drive field can break the
pair of total momentum $\qq$ into unpaired fermions of momenta $\kk+\qq/2$ and
$\kk-\qq/2$.
As a function of the increasing drive frequency $\omega$, this resonance condition is $(i)$ satisfied
by no wavevector when $\omega<\omega_1$ (with $\omega_1=2\Delta$ at low-$q$),
$(ii)$ satisfied for a connected set of wavevectors around the dispersion minimum
of the quasiparticle branch for  $\omega_1<\omega<\omega_2$, $(iii)$ satisfied by two connected sets of wavevectors,
one in the increasing and one in the decreasing part of the BCS branch for $\omega_2<\omega<\omega_3$
and $(iv)$ satisfied for a connected set of wavevectors in the increasing part of the branch for $\omega>\omega_3$.
These three boundary frequencies $\omega_1$, $\omega_2$ and $\omega_3$ will appear as angular points
in the spectral function and therefore in the response functions.
The quasiparticle-quasihole part of the spectral density (which originates from the second matrix in \eqref{Pi} with 
$\omega-(\epsilon_{\kk+\qq/2}-\epsilon_{\kk-\qq/2})$ in the denominator) is nonzero
when the ``absorption-emission'' resonance condition is satisfied
\be
\omega=\epsilon_{\kk+\qq/2}-\epsilon_{\kk-\qq/2}.
\ee
In this case, the drive field is not breaking the Cooper pairs
but simply transferring more energy to the unpaired fermions 
already created by thermal agitation. This requires much less energy,
which is why the quasiparticle-quasihole continuum is gapless. 
As a function of $\omega$, this resonance condition
can be met on $(i)$ two disconnected sets of wavevectors,
one in the increasing and one in the decreasing part of the BCS branch
for $\omega<\omega_{\rm ph}$, $(ii)$ a single connected set of wavevectors
 in the increasing part of the branch for $\omega>\omega_{\rm ph}$.
With $\omega_{\rm ph}$, we have found the four angular
point of the response function $\omega\mapsto\chi(\omega)$ in $[0,+\infty[$.
 
\res{
\section{Long wavelength limit}
\label{sec:petitsq}
In the long wavelength limit ($q\to0$) the solutions of the collective mode equation \eqref{modesco}
and the behavior of the response functions can be studied analytically.
Below the pair-breaking continuum (at energies lower than $\min_{\kk}(\epsilon_{\kk+\qq/2}+\epsilon_{\kk-\qq/2})$)
the problem has been studied in-depth at zero and nonzero temperature. At zero temperature
a real solution of \eqref{modesco} corresponding 
to the Anderson-Bogoliubov sound branch can be found \cite{Anderson1958,Strinati1998,CKS2006}. 
This branch appears as a pole in $\textrm{Re}\,\chi$ and as a Dirac peak in $\textrm{Im}\,\chi$ because
the zero-temperature response functions are free of branch cuts below the pair-breaking continuum.
This resonance was observed experimentally in the density-density response function 
by Bragg spectroscopy \cite{Vale2017} and in the low-$q$ limit it can be identified to hydrodynamic first sound \cite{Thomas2007,Stringari2013,Zwierlein2019}.
When the dispersion is supersonic (\textit{i.e.} the function $q\mapsto\omega_q$ is convex) 
the resonance is broadened by intrinsic effects not captured by RPA \cite{Beliaev1958,annalen}.

By contrast, the behavior of the response functions at high energy
($\omega>\min_{\kk}(\epsilon_{\kk+\qq/2}+\epsilon_{\kk-\qq/2})$)
is known in literature only at zero temperature. There, the collective mode equation \eqref{modesco} has a complex root
departing quadratically with $q$ from $2\Delta$ \cite{Popov1976}, and the
modulus-modulus response shows correspondingly a resonance
at energies above $2\Delta$ \cite{higgs}. Here, we show analytically
that (within the RPA) the same resonance exists at nonzero temperature, even
until the vicinity of $T_c$. This result is somewhat in disagreement
with Popov and Andrianov, who found no complex root close to $2\Delta$
in the limit $T\to T_c$. 

\subsection{General case}
We first compute the matrix elements $\Pi_{ij}$ 
on the upper-half complex plane, that is for $\textrm{Im}\, z>0$.
The long wavelength expansion can be perform
using the method exposed in \cite{higgs}: since the 
energy is expected to depart quadratically from $2\Delta$,
one parametrizes it as
\be
z=2\Delta+\zeta \frac{\mu}{\Delta}\frac{ q^2}{2m}.
\label{z2delta}
\ee
The window $[2\Delta,\omega_2]$ between the first two angular points of the
branch cut corresponds in the limit $q\to 0$ to $\zeta\in[0,1]$. With respect to the zero
temperature case \cite{higgs}, the phase-phase and modulus-modulus matrix elements
are simply multiplied by a factor $\mbox{tanh}(\Delta/2T)$:
\bea
\check\Pi_{11}(z,\qq)&\underset{q\to0}{\sim}&-\frac{\pi^2}{\check q}\mbox{tanh}\bb{\frac{\Delta}{2T}} f_{11}(\zeta) \qquad \mbox{with} \qquad f_{11}(\zeta)={\mbox{ln}\frac{1+\sqrt{1-\zeta}}{\sqrt{\zeta}}+\frac{i\pi}{2}}, \label{Pi11}\\
\check\Pi_{22}(z,\qq)&\underset{q\to0}{\sim}&{\pi^2 \check q}\frac{\mu}{2\Delta}\mbox{tanh}\bb{\frac{\Delta}{2T}} f_{22}(\zeta) \qquad \mbox{with} \qquad f_{22}(\zeta)= {\sqrt{1-\zeta}-\zeta \mbox{ln}\frac{1+\sqrt{1-\zeta}}{\sqrt{\zeta}}-\frac{i\pi\zeta}{2}},  \label{Pi22}
\eea
where we have introduced the dimensionless quantities $\check q=q/\sqrt{2m\Delta}$ and
$\check \Pi_{ij} =\alpha (\Pi_{ij}-L^3/g_0)$ for $(i,j)=(1,1)$ or $(2,2)$ and $\check \Pi_{ij} = \alpha \Pi_{ij}$ for the other matrix elements,
the nondimensionalization factor\footnote{
We have replaced the sums over $\kk$ in (\ref{Sigma}--\ref{S}) by an integral
$\int d^3k/(2\pi/L)^3$. The dimensionless response functions $\check\chi_{ij}$ 
are accordingly multiplied by the appropriate power of $\alpha$: $\alpha^{-1}$ for $\chi_{11}$, $\chi_{22}$ and
$\chi_{12}$, $\alpha^0$ for $\chi_{13}$ and $\chi_{23}$ and $\alpha^1$ for $\chi_{33}$.} $\alpha$ being $\Delta (2\pi/L\sqrt{2m\Delta})^3 $. 
We have written the complex functions $f_{11}$ and $f_{22}$ of $\zeta$ in such a way that
their spectral density (their imaginary part in the limit $\textrm{Im}\, \zeta\to0^+$) is directly given by their last term.
The modulus-phase matrix element is independent of $\zeta$ (it can be approximated by its value in $q=0$, $z=2\Delta$)
but, unlike $\Pi_{11}$ and $\Pi_{22}$, depends on temperature through the two ratios $\Delta/T$ and $\mu/T$:
\be
\check\Pi_{12}(z,\qq)\underset{q\to0}{\sim}\check\Pi_{12}(2\Delta,0) 
= {\pi^2}\sqrt{\frac{\mu}{2\Delta}}\mbox{tanh} \bb{\frac{\Delta}{2T}}g_{12}\bb{\frac{\mu}{T},\frac{\Delta}{T}}, \label{Pi12}
\ee
where we have introduced the improper integral\footnote{$\mathcal{P}$ denotes the Cauchy principal part)}:
\be
{\pi^2}\sqrt{\frac{M}{2D}}\mbox{tanh} \bb{\frac{D}{2}} g_{12}(M,D)=-\pi\mathcal{P}\int_{-M/D}^{\infty} \frac{d\xi\sqrt{\xi+{M}/{D}}}{\xi \sqrt{\xi^2+1}} \mbox{tanh} \frac{D\sqrt{\xi^2+1}}{2}.
\ee
Note that the quasiparticle-quasihole integrals \eqref{S} have been negligible
in deriving expressions (\ref{Pi11}--\ref{Pi12}).

To find a root of the collective mode equation \eqref{modesco},
one analytically continues $\Pi_{11}$ and $\Pi_{22}$ (and hence the determinant of $\Pi$) 
from upper to lower half-complex plane (the forms given in Eqs.~\eqref{Pi11}--\eqref{Pi22} are in fact already
analytic for $\mbox{Re}\,\zeta\in[0,1]$). The equation \eqref{modesco} for the complex collective mode frequency
$z_q=2\Delta+\zeta_s \frac{\mu}{\Delta}\frac{ q^2}{2m}+O(q^3)$ becomes an explicit (yet transcendental)
equation for the reduced frequency $\zeta_s$
\be
f_{11}(\zeta_s) f_{22}(\zeta_s)+g_{12}^2\bb{\frac{\mu}{T},\frac{\Delta}{T}}=0.
\label{eqzeta}
\ee
At zero temperature, this equation was derived in \cite{Popov1976} in the weak-coupling limit
and \cite{higgs} in the general case.
In this low-$q$ limit, the only dependence on temperature is
through the $\zeta$-independent second term of \eqref{eqzeta}.

\subsection{Close to the phase transition}
Unlike in the phononic regime ($q\to0$ with $z=cq$) \cite{artlongsk}, 
no dramatic phenomenon occurs for the collective mode of the pair-breaking continuum
when $T$ tends to the critical temperature $T_c$. We recall that in the RPA, the limit of the
phase transition from the superfluid phase corresponds to
\bea
\frac{\Delta}{T} &\to& 0 \label{limTc}\\
\frac{\mu}{T} &=& \frac{\mu(T_c)}{T_c} +O\bb{\frac{\Delta}{T}}^{2}. \label{limTc2}
\eea
The RPA also assumes an infinite fermionic quasiparticle lifetime
and thus describes the collective modes and their damping by the fermionic continua
in the collisionless approximation.

The function $g_{12}$ tends to
a finite nonzero constant in the limit $T\to T_c$:
\be
g_{12}\bb{\frac{\mu}{T},\frac{\Delta}{T}}\underset{T\to T_c}{\to}g_{12,c}
\bb{\frac{\mu_c}{T_c}}\equiv-\frac{{2}}{\pi}\sqrt{\frac{2T_c}{\mu_c}}\mathcal{P}\int_{-\mu_c/T_c}^{+\infty}\frac{\mbox{tanh}\frac{|e|}{2}\sqrt{e+\frac{\mu_c}{T_c}}de}{e|e|},
\ee
where we denote $\mu_c\equiv\mu(T_c)$.
In fact, the resonance near $T_c$ for a given value of $1/k_{\rm F}a$
has exactly the same shape as the $T=0$ resonance for a lower value (corresponding to weaker-coupling) 
of $1/k_{\rm F}a$. Using an equation-of-state to relate $1/k_{\rm F}a$
to both $\mu_c/T_c$ and $\mu(T=0)/\Delta(T=0)$ \cite{artlongsk}, 
the corresponding values $a_0$ and $a_c$ of the scattering length
at $T=0$ and $T_c$ are found by solving:
\be
g_{12,c}\bb{\frac{\mu_c}{T_c}\Big\vert_{a=a_c}}=\lim_{T\to0}g_{12}\bb{\frac{\mu}{\Delta}\Big\vert_{a=a_0}\frac{\Delta}{T},\frac{\Delta}{T}}.
\label{corresp}
\ee
Finally, the explicit expressions of the response functions in the long wavelength limit, at arbitrary $0\leq T<T_c$, and in the limit $T\to T_c$ are:
\bea
\!\!\!\!\!\!\!\!\!\chi_{11}(z,\qq) &=&-\frac{\check q}{\pi^2 \mbox{tanh}\frac{\Delta}{2T}} \frac{f_{22}(\zeta)}{f_{11}(\zeta)f_{22}(\zeta) + g_{12}^2(\mu/T,\Delta/T)} +O(q^2) \underset{T\to T_c}{\sim} -\frac{2\check q}{\pi^2}\frac{T_c}{\Delta}  \frac{f_{22}(\zeta)}{f_{11}(\zeta)f_{22}(\zeta) + g_{12,c}^2(\mu_c/T_c)} \\ 
\!\!\!\!\!\!\!\!\!\chi_{22}(z,\qq) &=&\frac{1}{\check q \pi^2  \mbox{tanh}\frac{\Delta}{2T}} \frac{2\Delta}{\mu} \frac{f_{11}(\zeta)}{f_{11}(\zeta)f_{22}(\zeta) + g_{12}^2(\mu/T,\Delta/T)} +O(1) \underset{T\to T_c}{\sim} \frac{4}{\pi^2 \check q} \frac{T_c}{\mu_c} \frac{f_{11}(\zeta)}{f_{11}(\zeta)f_{22}(\zeta) + g_{12,c}^2(\mu_c/T_c)}\\
\!\!\!\!\!\!\!\!\!\chi_{12}(z,\qq) &=&-\frac{1}{\pi^2  \mbox{tanh}\frac{\Delta}{2T}} \sqrt{\frac{2\Delta}{\mu}} \frac{g_{12}(\mu/T,\Delta/T)}{f_{11}(\zeta)f_{22}(\zeta) + g_{12}^2(\mu/T,\Delta/T)} +O(q) \!\!\! \underset{T\to T_c}{\sim} \!\!\! -\frac{2\sqrt{2}}{\pi^2  }\frac{T_c}{\sqrt{\mu_c \Delta}}  \frac{g_{12,c}(\mu_c/T_c)}{f_{11}(\zeta)f_{22}(\zeta) + g_{12,c}^2(\mu_c/T_c)}.
\eea
Thus, the response functions have exactly the same shape (they coincide up to a proportionality factor)
near $T_c$ than at $T=0$ for the slightly different value of the interaction strength given by \eqref{corresp}. 

In Fig.~\ref{fig:crossover}, we show how the shape\footnote{Note that the response functions are rescaled by the power of $q$
ensuring a finite non zero limit when $q\to0$. Similarly the response functions near $T_c$ are rescaled by the right power of $\Delta/T_c$. Finally, a proportionality factor (depending of $\mu/\Delta$ at $T=0$ and $\mu_c/T_c$ near $T_c$) is applied to ensure that the response functions at $T=0$ and $T\to T_c$ fall on the same limit.} of the order-parameter response functions ($\chi_{11}$, $\chi_{22}$
and $\chi_{12}$) change when going from the BCS limit ($1/k_F a\to -\infty$ that is $\mu/\Delta\to+\infty$ at $T=0$ or $\mu_c/T_c\to+\infty$ at $T=T_c$) to the threshold of the BEC regime where $\mu$ vanishes. Exploiting the equivalence \eqref{corresp}, the figure describes together the crossover at $T=0$ and $T\to T_c$. Irrespectively of the interaction regime, the phase-phase response is a monotonously increasing function of the drive frequency and only reflects the incoherent response of the pair-breaking continuum,
without collective effects.
Conversely, both the modulus-modulus and modulus-phase response  
functions display a maximum that can be interpreted as a collective mode 
in the BCS limit (black curves) and up until unitarity (blue curves). As explained in Ref.~\cite{higgs},
this maximum can be fitted to extract the frequency and damping rate of the collective mode to a good precision. The fit function to use is $\omega\mapsto \mbox{Im} (Z_q/(\omega-z_q)+C_q)$, where the complex parameters $z_q$, $Z_q$ and $C_q$ represent respectively the complex energy of the collective mode, its residue, and an incoherent flat background. A
remarkable effect of this background $C_q$ is to displace 
the location of the maximum of $\chi_{22}$ and $\chi_{12}$ to respectively $\zeta\simeq0.4$ and $\zeta\simeq0.1$
in a very broad interaction range. The variations of the real part of the root $\zeta_s$
(which decreases when increases the coupling strength) are thus not visible by simply looking at the
maximum location. 
Soon after unitarity, the resonance in $\chi_{22}$ and $\chi_{12}$ disappears and only a sharp feature near $\omega=2\Delta$ remains.
This abrupt lower edge of the continuum is in $\zeta=0$ so it is not departing quadratically with $q$ 
from $2\Delta$ (see also the color figure \ref{fig:couleurBEC} in the BEC regime) 
as the Popov-Andrianov resonance does in the BCS regime, 
and it can no longer be interpreted as a collective mode.
As understood in \cite{higgs}, this is because the complex root $z_q$ of the collective mode equation \eqref{modesco}
has a real part below $2\Delta$ (i.e $\textrm{Re}\, \zeta_s<0$) and does no longer trigger a resonance inside the pair-breaking continuum.

\begin{figure}[htb]
\includegraphics[width=0.49\textwidth]{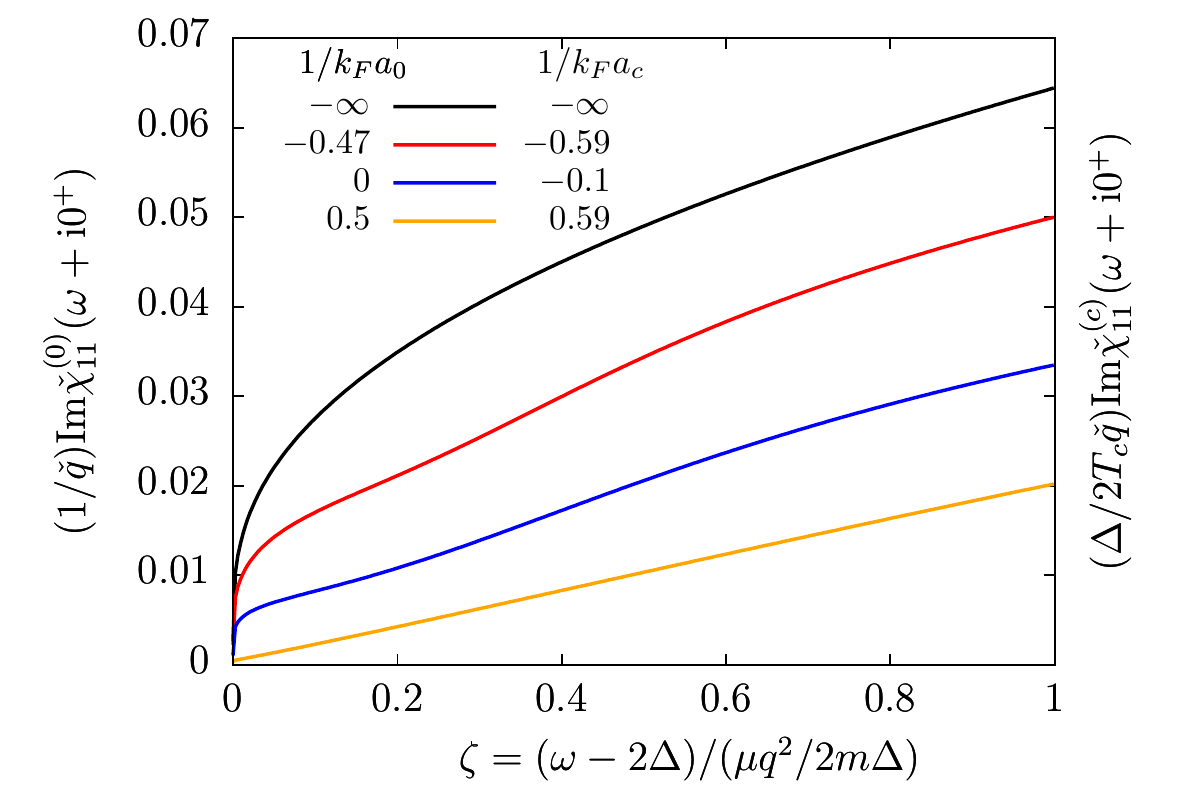}
\includegraphics[width=0.49\textwidth]{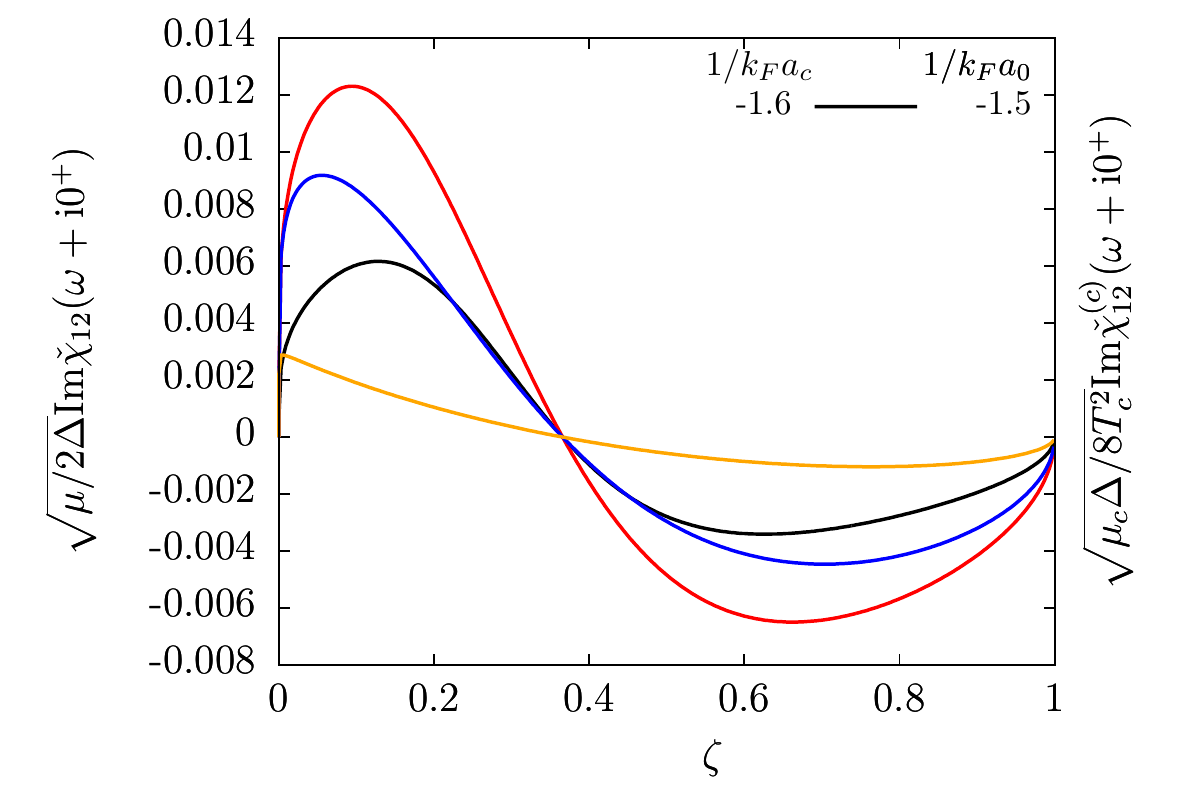}

\includegraphics[width=0.6\textwidth]{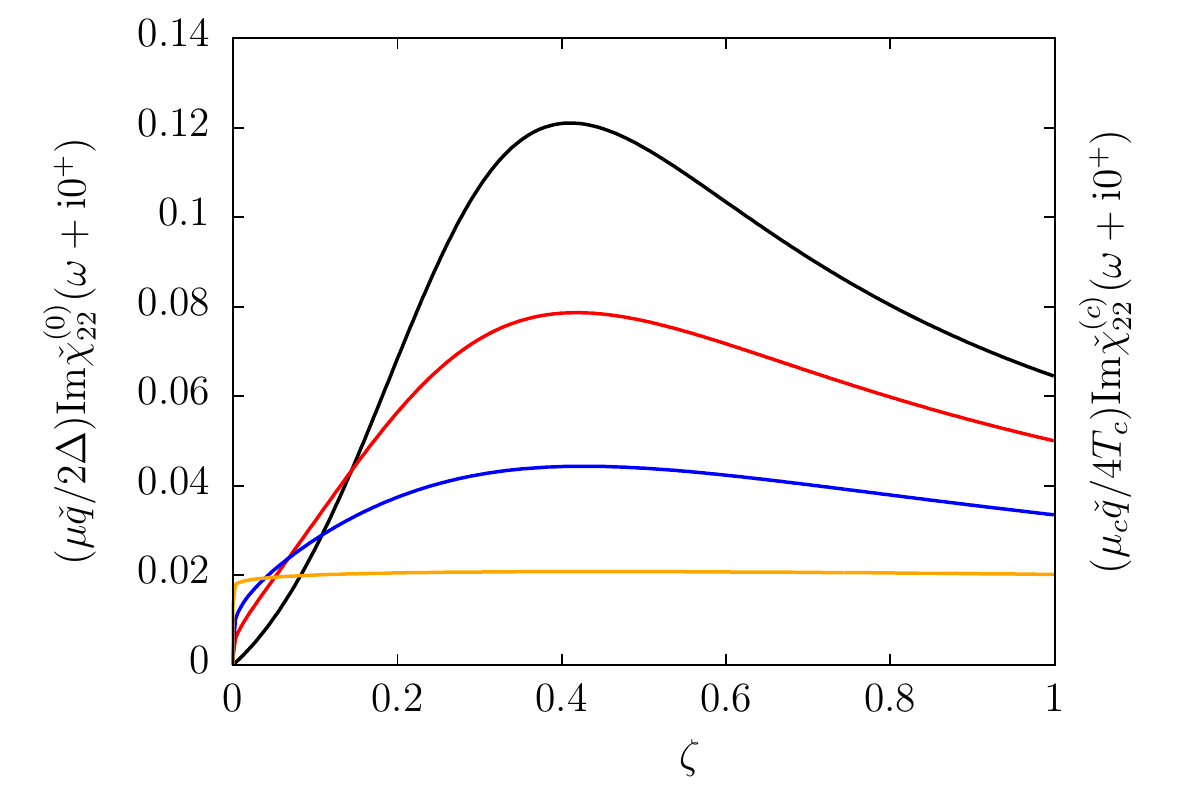}

\caption{\label{fig:crossover} The order-parameter response functions (top left pannel: phase-phase, top right pannel: modulus-phase, bottom pannel: modulus-modulus response) are shown as functions of the reduced drive frequency $\zeta=(\omega-2\Delta)/(\mu{q}^2/2m\Delta)$ of Eq.~\eqref{z2delta} in the long wavelength limit after multiplication by the power of $q$ which ensures a finite non zero limit when $q\to0$. Their value at zero temperature ($\chi^{(0)}_{ij}$) coincide up to a proportionality factor (shown in the $y$-axis labels) with their value near the phase transition ($\chi^{(c)}_{ij}$) at a slightly different interaction strength $1/k_F a$ obtain using the correspondence Eq.~\eqref{corresp}. The values of $1/k_F a$ used to span the BCS side of the crossover at $T=0$ correspond with the mean-field equation-of-state to $\mu/\Delta=+\infty$, $2$, $0.86$ and $0.07$ (respectively black, blue, red and orange lines). At $T\to T_c$ they correspond to $\mu_c/T_c=+\infty$, $4.2$, $1.8$ and $0.15$. For $\chi_{12}$, the BCS limit $\mu/\Delta,\mu_c/T_c\to+\infty$ is reached logarithmically such that we have used the finite values $\mu/\Delta=10$, $1/k_F a_0=-1.5$ at $T=0$ and $\mu_c/T_c=20.6$, $1/k_F a_{c}=-1.6$ near $T_c$.}
\end{figure}

\subsection{Density matrix elements in the long wavelength limit}
We now study the density responses of the system $\chi_{i3}$, $i=1,2,3$ in the 
long wavelength limit $q\to0$ at energies above but close to $2\Delta$.
In the long wavelength limit $q\to0$ (at fixed $\zeta$), the three
matrix elements needed to compute the density response functions are given by:
\bea
\delta\check\Pi_{13}&\equiv& \check\Pi_{13}+\check\Pi_{11}={\pi^2}\frac{\mu}{2\Delta}  \mbox{tanh}\bb{\frac{\Delta}{2T} } f_{13}(\zeta) \check q +O(q^2)  \qquad \mbox{with} \qquad f_{13}(\zeta)=\sqrt{1-\zeta}\ \label{Pi13}\\
\delta \check\Pi_{23}&\equiv& \check\Pi_{23}+\check\Pi_{12}=\pi^2 \bb{\frac{\mu}{2\Delta}}^{3/2}  \mbox{tanh}\bb{\frac{\Delta}{2T} } \bbcro{\zeta g_{12}\bb{\frac{\mu}{T},\frac{\Delta}{T}} + g_{23}\bb{\frac{\mu}{T},\frac{\Delta}{T}}}\check q^2+O(q^3)\\
\delta\check\Pi_{33}&\equiv& \check\Pi_{33}+\check\Pi_{11}+2\check\Pi_{13}= g_{33}\bb{\frac{\mu}{T},\frac{\Delta}{T}}\check q^2+\frac{\pi^2}{2}\frac{\mu^2}{4\Delta^2}\mbox{tanh}\bb{\frac{\Delta}{2T}} f_{33}(\zeta)\check q^3+O(q^4) \label{Pi33}\\
\mbox{with} \qquad f_{33}&=& {(\zeta-2)\sqrt{1-\zeta}-\zeta^2\mbox{ln}\frac{1+\sqrt{1-\zeta}}{\sqrt{\zeta}}-\frac{i\pi\zeta^2}{2}} \\
   \pi^2 \bb{\frac{M}{2D}}^{3/2} && \!\!\!\!\!\!\!\!\!\!\!\! \mbox{tanh}\bb{\frac{D}{2} }  g_{23}(M,D)=\frac{\pi}{3} \mathcal{P}\int_{-M/D}^{\infty} \frac{d\xi\sqrt{\xi+{M}/{D}}^3}{\xi \epsilon^3} \mbox{tanh} \frac{D\epsilon}{2}+\frac{\pi D}{6}  \mathcal{P}\int_{-M/D}^{\infty} \frac{\xi d\xi\sqrt{\xi+{M}/{D}}^3}{ \epsilon^2 \mbox{cosh}^2 \frac{D\epsilon}{2} } \\
&&\mbox{and} \!\!\!\! \qquad  g_{33}(M,D)=\frac{\pi}{16} \mathcal{P}\int_{-M/D}^{\infty} \frac{d\xi\sqrt{\xi+{M}/{D}}}{\mbox{cosh}^2\frac{D\epsilon}{2} \epsilon^5} \bbcrol{-2D\epsilon\bb{\xi\epsilon^2+\frac{2}{3}\bbcro{1-2{\xi^2(1+\epsilon^2)}}\bb{\xi+{M}/{D}}}}\\
&&\qquad\qquad\qquad \bbcror{+\frac{\mbox{sinh}\frac{3D\epsilon}{2}}{\mbox{cosh}\frac{D\epsilon}{2}}\bb{\xi\epsilon^2+2\bb{\xi+{M}/{D}}}
+\mbox{tanh}\frac{D\epsilon}{2}\bb{\xi\epsilon^2+2\bb{\xi+{M}/{D}}\bb{1+\frac{2}{3}D^2\xi^2\epsilon^2} }}. \notag
\eea
Those expressions, like those of the modulus and phase matrix elements (\ref{Pi11}--\ref{Pi12})
are obtained by treating separately the resonant wavevectors (for the resonance condition $z=\epsilon_{\kk+\qq/2}+\epsilon_{\kk-\qq/2}$),
located in this limit around the minimum $k_0$ of the BCS branch. For those wavevectors, we set 
\be
k=k_0+Kq,
\label{K}
\ee 
and expand the integrand in \eqref{Sigma} at fixed $K$. This yields the leading order contribution to $\Pi_{11}$, $\Pi_{22}$ and $\delta\Pi_{13}$.
For $\Pi_{12}$, $\delta\Pi_{23}$ and $\delta\Pi_{33}$ the leading order
is dominated by the wavevectors away from $k_0$ and is obtained by expanding
directly in powers of $q$ at fixed $k$ (with a contribution of
the quasiparticle-quasihole integrals from Eq.~\eqref{S}).
For  $\delta\Pi_{33}$ specifically, the subleading order $O(q^3)$ (which matters 
for the imaginary part of the response function $\textrm{Im}\,\chi_{33}$), is obtained by subtracting the leading one
and then using the reparametrisation of the wavevectors, Eq.~\eqref{K}.

Using the expansions Eqs.~(\ref{Pi13}--\ref{Pi33}), we obtain the expressions
of the density response functions:
\bea
\chi_{13} &=& 1+\frac{\mu}{2\Delta} \frac{(\zeta g_{12}+g_{23})g_{12}-f_{13} f_{22}}{f_{11} f_{22}+g_{12}^2}
\check q^2 + O(q^3) \\
\chi_{23} &=& -\sqrt{\frac{\mu}{2\Delta}} \frac{f_{13} g_{12} + (\zeta g_{12}+g_{23})f_{11}}{f_{11} f_{22}+g_{12}^2} \check q + O(q^2) \\
\chi_{33} &=& g_{33}\check q^2 +\pi^2\frac{\mu^2}{4\Delta^2} \mbox{tanh}\bb{\frac{\Delta}{2T} } \bbcro{\frac{f_{33}}{2}-\frac{2f_{13}(\zeta g_{12}+g_{23})g_{12}-f_{13}^2f_{22}+(\zeta g_{12}+g_{23})^2f_{11}}{f_{11} f_{22}+g_{12}^2}}\check q^3+O(q^4),
\eea
where we omit the evaluation of the functions $f_{ij}$ and $g_{ij}$ respectively in $\zeta$ and $(\mu/T,\Delta/T)$. The limiting behaviour
near the phase transition follows immediately by using the limiting behaviour of $\Delta$ and $\mu$
from Eqs.~(\ref{limTc}--\ref{limTc2}) and replacing $g_{12}(\mu/T,\Delta/T)$ and $g_{23}(\mu/T,\Delta/T)$ by their finite
nonzero limit $g_{12,c}(\mu/T_c)$ and $g_{23,c}(\mu/T_c)$ with
\be
g_{23,c}(M)= \frac{2\sqrt{2}}{3\pi M^{3/2}} \int_{-M}^{+\infty} \frac{de (e+M)^{3/2}}{e\mbox{cosh}^2 (e/2)}.
\ee
The function $g_{33}$, which gives only a $\zeta$-independent shift of $\mbox{Re}\, \chi_{33}$,
diverges asymptotically as $O(\Delta/T_c)^{-3/2}$ near the phase transition.

In Fig.~\ref{fig:crossoverTc}, we show the density response functions
on the BCS side of the crossover. Unlike for the order-parameter response
functions, no exact correspondance between zero temperature and the transition temperature can
be found by changing the interaction strength (this is due to the
temperature dependence of $g_{23}$), so we show separately
the functions at $T=0$ (in solid curves) and at $T\to T_c$ (in dashed curves).
The difference  between the $T=0$ and $T\to T_c$ curves (after the appropriate rescaling)
remains however fairly small, and tends to 0 in the BCS limit (black curves).
Remarkably, a minimum characteristic of the Popov-Andrianov
collective mode is visible in all three density responses. In $\chi_{23}$, this
minimum is a global minimum (for $\zeta\in[0,1]$) which exist (as in $\chi_{22}$ and  $\chi_{12}$) 
from the BCS limit up until unitarity. For the density-density and density-phase responses $\chi_{33}$
and $\chi_{13}$, this minimum is a local minimum, which exists close to unitarity (blue curves)
around $\zeta=0.1$. Because of the decoupling between the phase-density fluctuations
and the modulus fluctuations in the weak-coupling limit,
this minimum disappears from $\chi_{13}$ and $\chi_{33}$ in the weak-coupling limit $1/k_F a\to-\infty$ (black curves).
After unitary, when approaching the BEC regime (orange curves),
the resonances in all three density responses are replaced
by a sharp edge in $\omega\to2\Delta^+$ ($\zeta\to0^+$).
This is the same phenomenon as in the order-parameter response functions.

\begin{figure}[htb]
\includegraphics[width=0.49\textwidth]{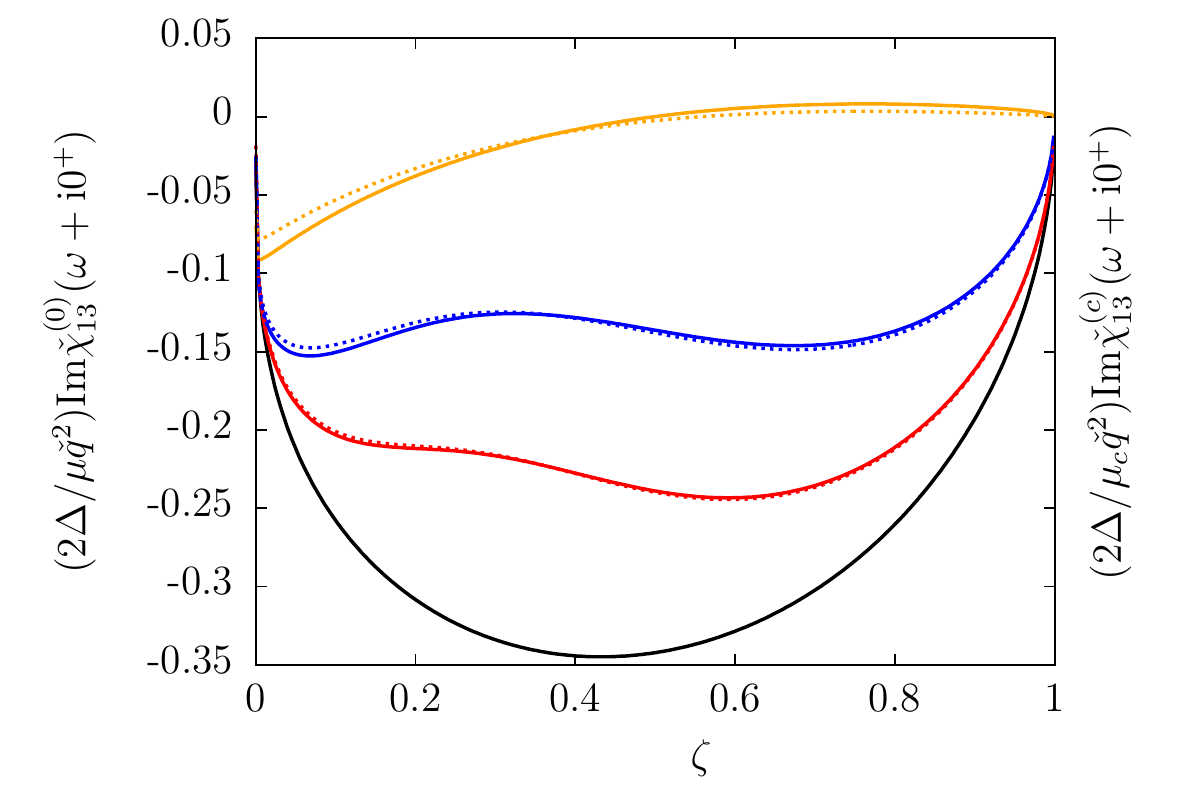}
\includegraphics[width=0.49\textwidth]{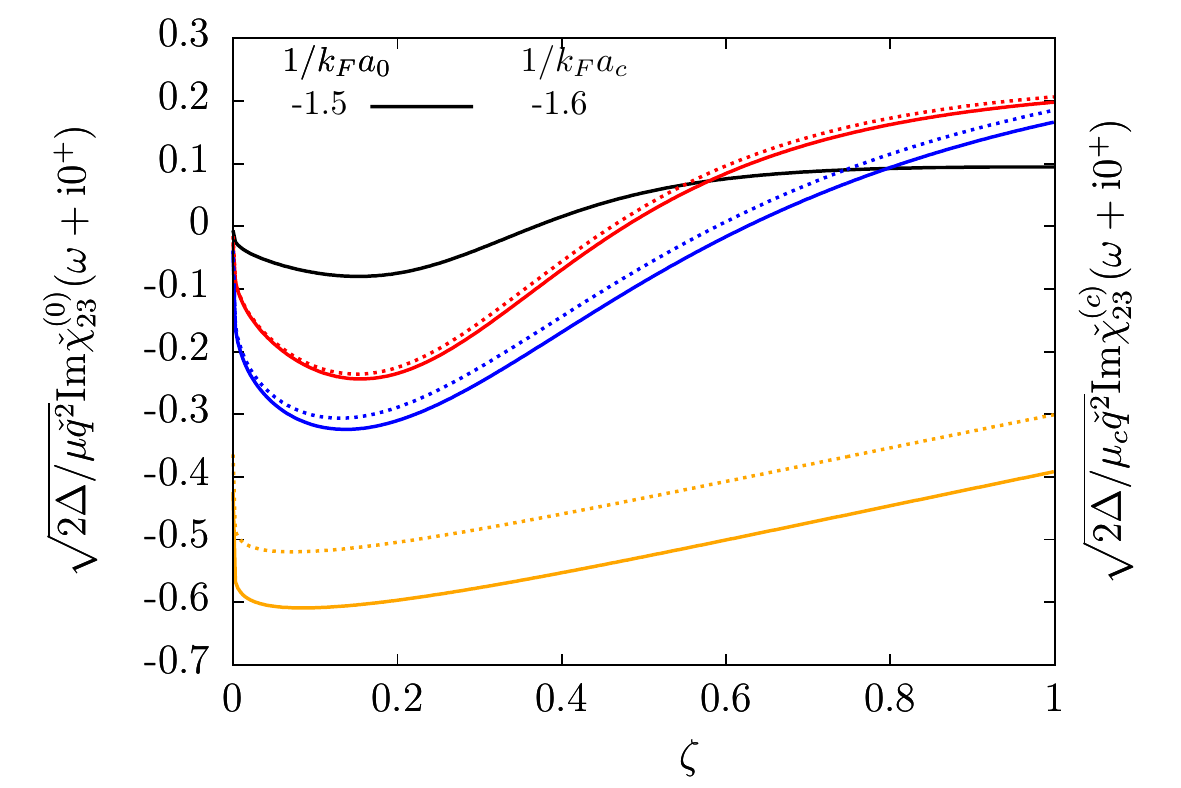}

\includegraphics[width=0.6\textwidth]{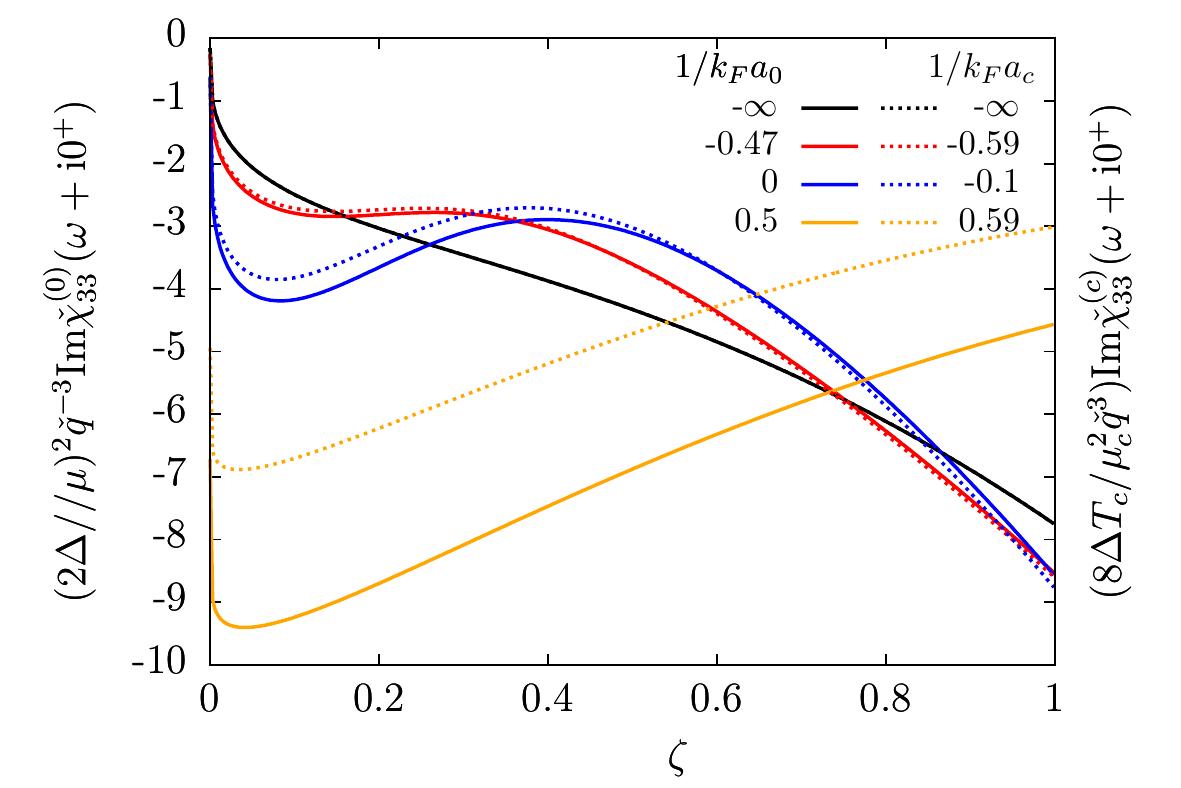}
\caption{\label{fig:crossoverTc} The density response functions (top left pannel: density-phase, top right pannel: density-modulus, bottom pannel: density-density response) are shown as functions of the reduced drive frequency $\zeta=(\omega-2\Delta)/(\mu{q}^2/2m\Delta)$ of Eq.~\eqref{z2delta} in the long wavelength limit after multiplication by the power of $q$ which ensures a finite non zero limit when $q\to0$. Their values at zero temperature ($\chi^{(0)}_{ij}$, solid lines) are compared to their value near the phase transition ($\chi^{(c)}_{ij}$, dashed lines) after the appropriate rescaling and the change of interaction strength which brings the order-parameter responses on the same line (see Fig.~\ref{fig:crossover}, and the correspondence Eq.~\eqref{corresp}). Please refer to the caption of Fig.~\ref{fig:crossover} for the values of $\mu/\Delta$ and of $\mu_c/T_c$ corresponding to the chosen values of $1/k_F a$.}
\end{figure}

\subsection{Coexistence with the phononic collective modes near $T_c$}

To compare the Popov-Andrianov resonance to the other collective effects of a superfluid Fermi gases near $T_c$,
we show in Fig.~\ref{fig:presdeTc2} the response functions from $\omega=0$ up until $\omega>3\Delta$ in the strong coupling
regime and temperature close to $T_c$.
The sharpest feature in both the order-parameter and density responses is the resonance, at very low energy (that is at $\omega=uq$ with a velocity $u\propto\sqrt{T_c-T}$), of the collisionless phononic collective mode found in \cite{artlongsk}.
Still at phononic energies $\omega\propto q$, the density-density response function shows a broad peak
caused entirely by $\Pi_{33}$ (shown as a black dashed line) and also noticed in \cite{artlongsk}. 
This peak exists also in the normal phase and may be interpreted as the zero sound of an ideal Fermi gas.
Finally, inside the first window $[2\Delta,\omega_2]$ of analyticity of the pair-breaking continuum, all response functions
show the peak characteristic of the Popov-Andrianov resonance, whose shape matches the one shown on 
Figs.~\ref{fig:crossover} and \ref{fig:crossoverTc}. Due to the absence of rescaling with the wavevector $q$ in Fig.~\ref{fig:presdeTc2}, the peak is much more
intense in the modulus-modulus response, and to a lesser extent in the modulus-density response, than in the density-density response.

\begin{figure}[htb]
\includegraphics[width=0.9\textwidth]{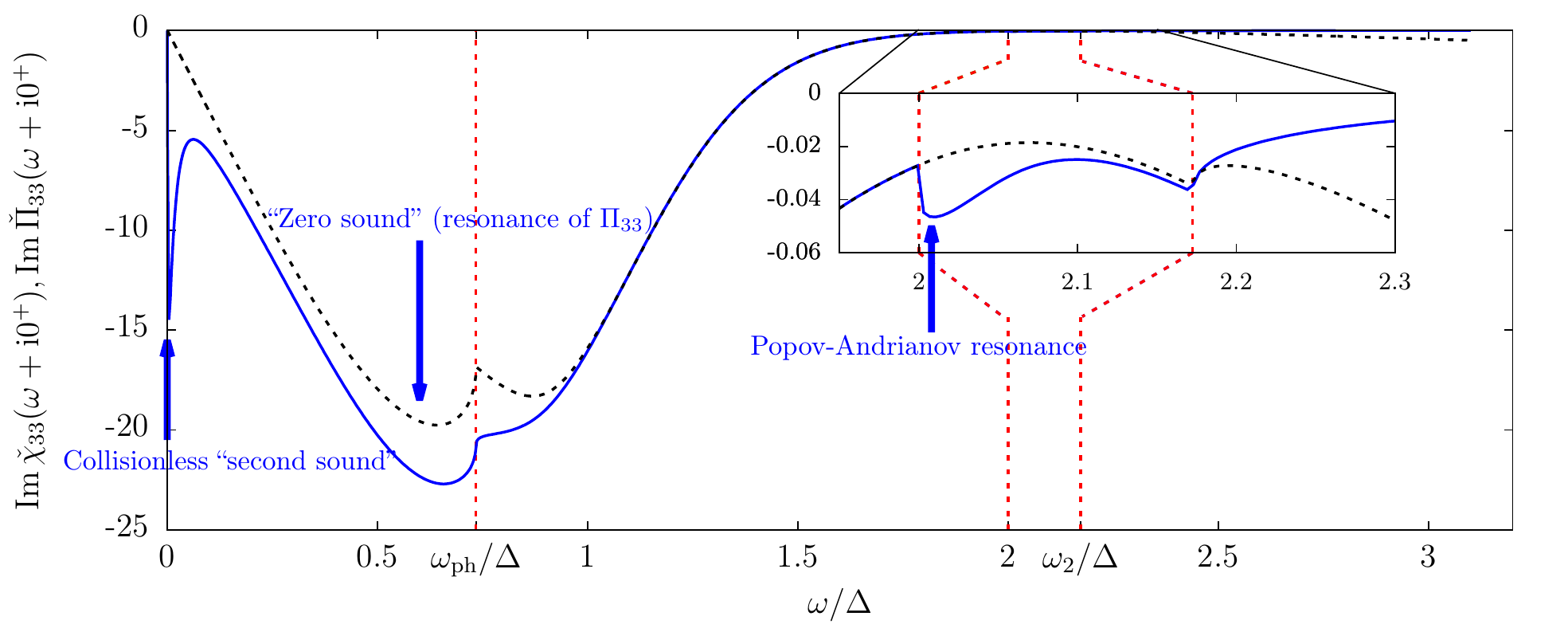}

\includegraphics[width=0.49\textwidth]{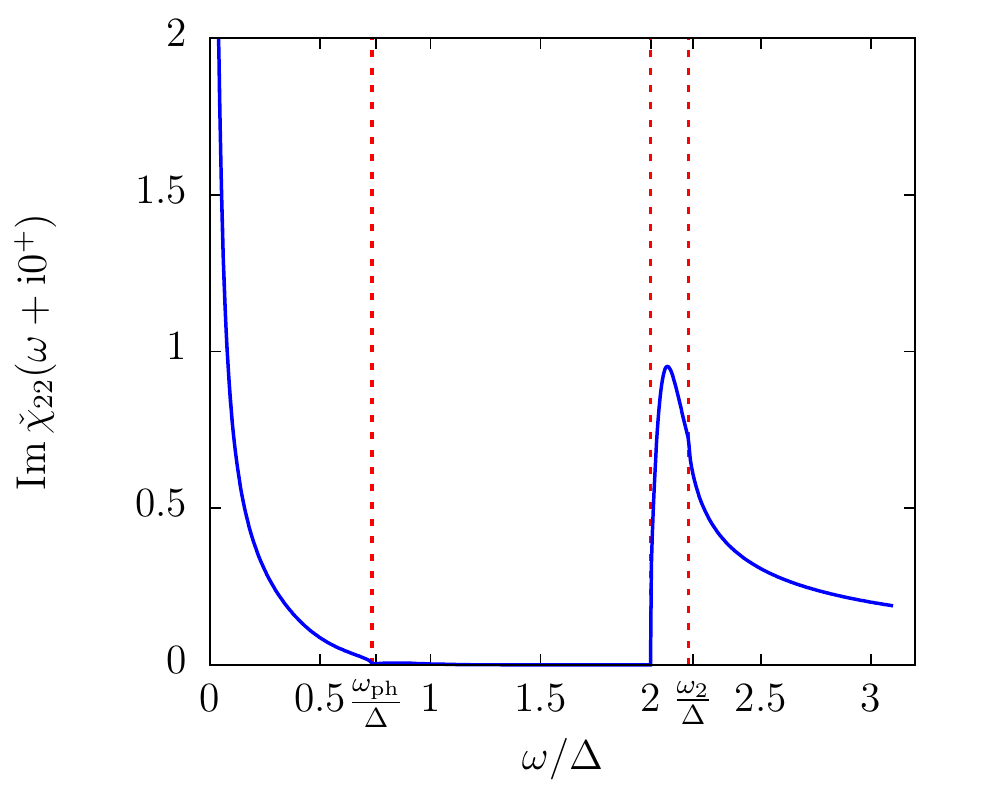}
\includegraphics[width=0.49\textwidth]{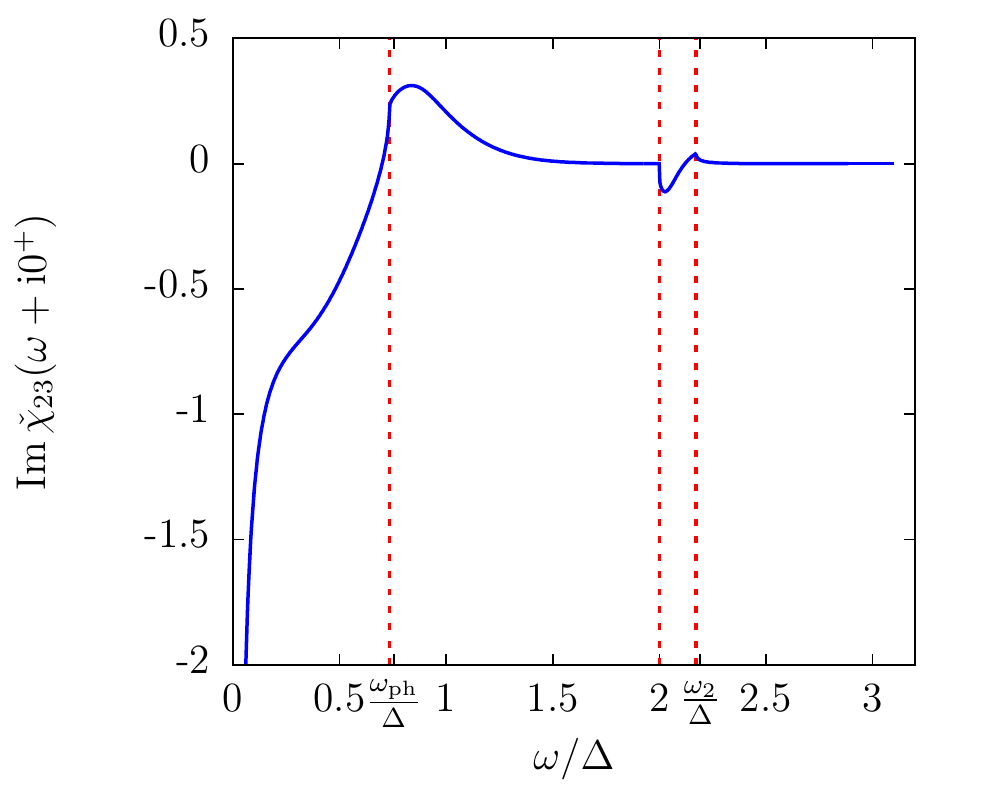}
\caption{\label{fig:presdeTc2} The density-density (top pannel), modulus-modulus (bottom left pannel) and modulus-density (bottom right pannel) response functions shown in function of the drive frequency $\omega$. The curves were drawn in the strong coupling regime ($1/k_{\rm F}a\simeq-0.1$, $\mu_c/T_c\simeq1.8$), near the transition temperature ($\Delta/T=0.1$) and at long wavelength ($\check q=0.1$). The dimensionless response functions are shown this time without rescaling and from $\omega=0$ until far inside the pair-breaking continuum. The vertical dashed lines show the angular points of the fermionic continua, from left to right $\omega_{\rm ph}$, $\omega_1$($=2\Delta$) and $\omega_2$. On the top panel, the black dashed line is the pure density contribution $\mbox{Im}\ \Pi_{33}$ to the density-density response (see Eq.~\eqref{chi33}) and the inset is a zoom on the behaviour near $\omega=2\Delta$.}
\end{figure}

\subsection{Experimental protocol}

Our results suggest a very simple experimental protocol to observe the resonance: 
using a Bragg spectroscopic measurement as in Ref.~\cite{Vale2017}, one should
observe that the first extremum above $2\Delta$ varies quadratically (both in location and width)
with $q$, a behavior which can be viewed as the fingerprint of the Popov-Andrianov-Higgs mode.
The optimal interaction regime is around unitarity and the optimal wavevector is around $0.5\times\sqrt{2m\Delta}$
($q$ should not be too small to avoid the $q^3$ cancellation of $\chi_{33}$ near $2\Delta$ but not too large either
otherwise the minimum is reabsorbed by the continuum edge, see the lower panels of Fig.~\ref{fig:LU}).
 
Alternatively, the resonance could be observed through the
modulus-density response function $\chi_{23}$
by $(i)$ exciting the order-parameter modulus $\delta|\Delta|$ through a 
modulation of the scattering length at frequency $\omega$ and 
wavelength $2\pi/q$ and $(ii)$ measuring the intensity of the density 
modulation $\delta\rho$ at wavelength $2\pi/q$. This should be
easier than the scheme of Ref.~\cite{higgs} which proposed to measure
 $\chi_{22}$ by interferometry. Using the 
 symmetry of the response matrix $\chi$, one can also excite
 the density $\delta\rho$ (by a Bragg pulse \cite{Vale2017} 
 or using the trapping potential \cite{Zwierlein2019})
and measure the order-parameter modulus $\delta|\Delta|$ 
either by interferometry or by bosonizing the Cooper pairs
through a fast sweep of the scattering length, as was done in \cite{Koehl2018}.

\section{At shorter wavelengths}

Outside the long wavelength limit, that is\footnote{see Eq.~(90) in \cite{higgslong} for a more detailed discussion the limit of validity of the long-wavelength limit} 
$q\approx\sqrt{2m\Delta},\sqrt{2m\mu}$ when $\mu$ and $\Delta$ are comparable,
we study the response functions by
performing numerically the integral over internal wavevectors $\kk$ in
Eqs.~\eqref{S} and \eqref{Sigma} (see Appendix \ref{app:numerique}
for more details on the numerical implementation).

\subsection{At zero temperature}

\subsubsection{Weak-coupling regime}

\begin{figure}[htb]
 \begin{minipage}{\textwidth}
\includegraphics[width=0.49\textwidth]{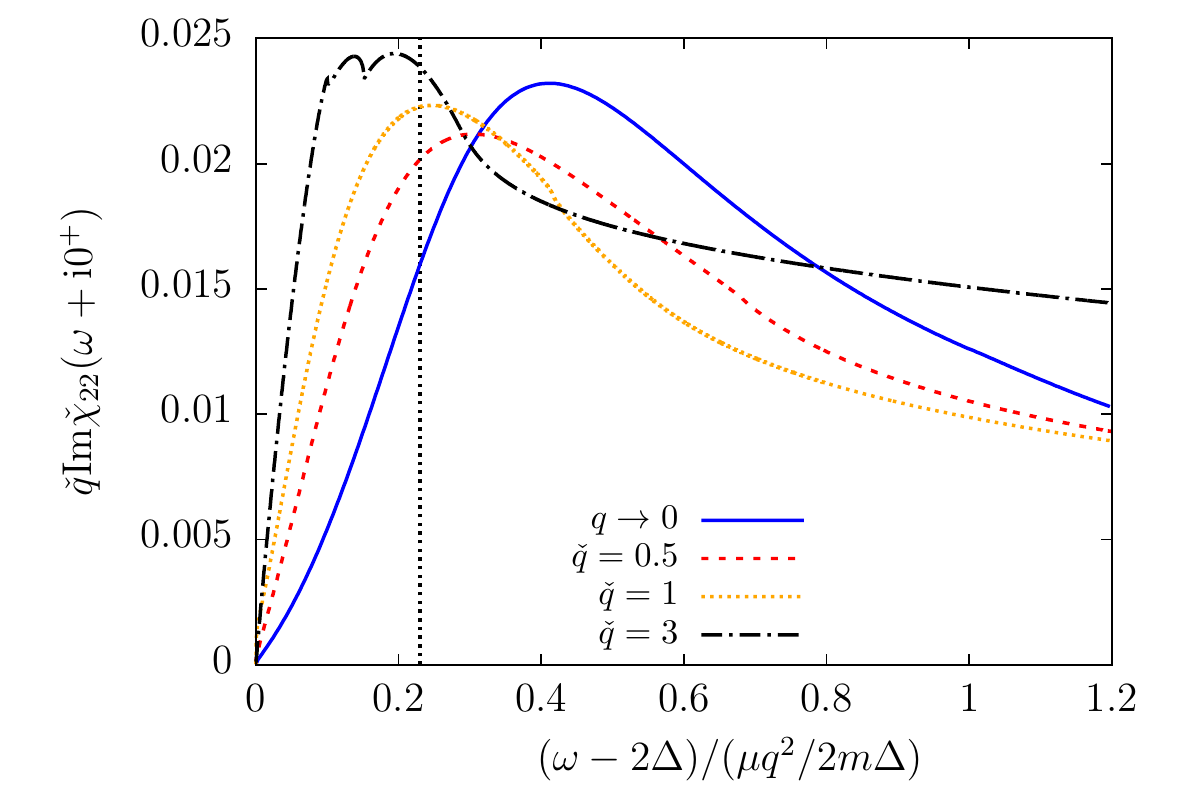}
\includegraphics[width=0.49\textwidth]{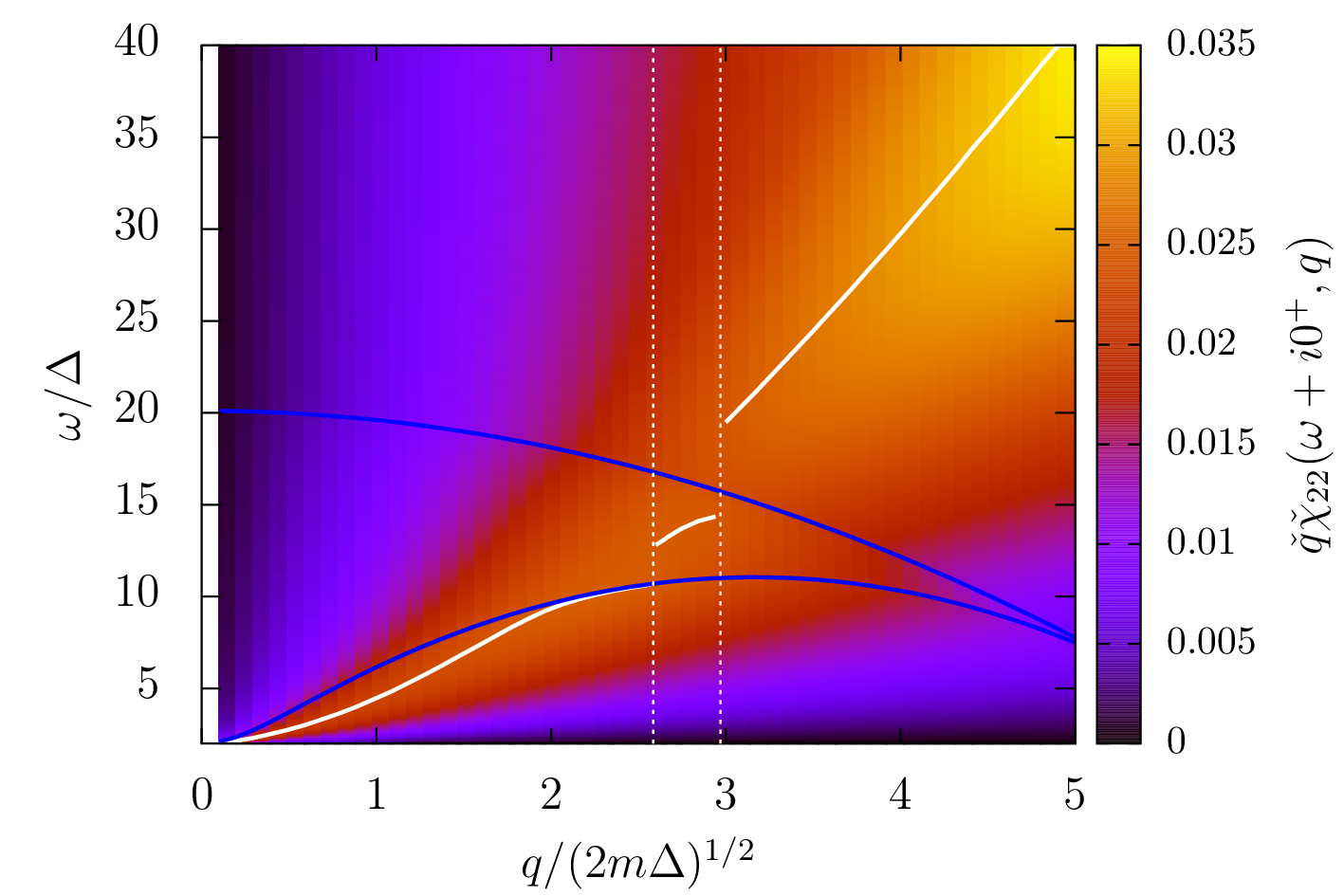}
\caption{\label{fig:T0} Left panel: the modulus-modulus response function is shown as a function of $(\omega-2\Delta)/(\mu{q}^2/2m\Delta)$ at weak-coupling $\mu/\Delta=10$. The wavevector $q$ varies from $q\to0$ (solid blue line, see section \ref{sec:petitsq}), $\check q=0.5$ (red dashed line), $\check q=1$ (orange dotted line) to $\check q=3$ (black dash-dotted line), and the response function is multiplied by $\check q$. The vertical dotted line indicates the reduced eigenergy $\mbox{Re}\zeta_s\simeq0.23$ of the pole found (when $q\to0$) in the analytic continuation (see Eq.~\ref{eqzeta}) through $[2\Delta,\omega_2]$. 
Right panel: the same evolution is shown in
colors as functions of both the wavevector $q$ on the $x$-axis and drive frequency $\omega$ on the $y$-axis. The response function is still multiplied by $\check q$. Superimposed to the color plot, the angular points $\omega_2$ (lower blue solid line) and $\omega_3$ (upper blue solid line), and the location of the global maximum of the function $\omega\mapsto\chi_{22}$ (white solid line). As $q$ increases, this maximum jumps from the interval $[2\Delta,\omega_2]$ where it is located at low $q$, to $[\omega_2,\omega_3]$ and eventually to $[\omega_3,+\infty[$ at large $q$. Each jump is marked by a vertical white dotted line.}
\end{minipage}
\end{figure}

On the left panel of Fig.~\ref{fig:T0}, we show the modulus-modulus response
at relatively weak-coupling ($\mu/\Delta=10$) and zero temperature
as a function of $\omega$ (rescaled as in the long wavelength section)
for increasing values of the wavevector $q$. On the right panel, we show
the same dispersion relation but in colors, with 	$q$ on the $x$-axis and
$\omega$ on the $y$-axis. The Popov-Andrianov resonance we have characterized
at low $q$ remains as a broader and shallower maximum as $q$ increases (see the rescaling of 
the $x$ and $y$-axis on the left panel of Fig.~\ref{fig:T0}) that travels roughly quadratically through the continuum.
In the modulus-modulus response function, the augmentation of wavevector is thus
unfavourable for the observation of the resonance in the pair-breaking continuum.
Note that the location of the maximum is discontinuous when crossing $\omega_2$ and $\omega_3$ (which both decrease with $q$),
but remains a monotonously increasing function of $q$. The non-monotonic behavior
of the collective mode eigenfrequency $z_q$ found in the analytic continuation through the
interval $[2\Delta,\omega_2]$ of the real axis \cite{higgs} is thus not reflected on the response function.
In fact the angular points $\omega_2$ and $\omega_3$ only slightly
affect the shape of the resonance when they cross it (see in particular the black curve on the
left panel of Fig.~\ref{fig:T0}). This is consistent with the finding
of Ref.~\cite{higgslong} (see in particular section 4.8 therein): 
at large $q$, the analytic continuations through windows 
$[\omega_2,\omega_3]$ and $[\omega_3,+\infty[$ predict 
a pole with an eigenfrequency close to that of the Popov-Andrianov  
branch in window $[2\Delta,\omega_2]$.
The same robustness towards the choice of the real axis interval 
through which the analytic continuation is made was noticed by Ref.~\cite{artlongsk}
for the phononic modes. \res{It is a sign that the Popov-Andrianov collective mode
is a fundamental physical phenomenon, which does not depend on a specific configuration of the 
fermionic continuum.}

\subsubsection{Strong-coupling regime}

Conversely, the increase of $q$ favours the observability of the resonance
in both the modulus-density and density-density response functions at strong coupling.
On Fig.~\ref{fig:LU}, we show $\chi_{23}$ and $\chi_{33}$ (as well as $\chi_{22}$) at
unitarity ($\mu/\Delta\simeq0.86$) and still at zero temperature. 
As long as it does not encounter the singularity in $\omega_3$,
a smooth extremum (in $\chi_{23}$ and $\chi_{33}$ it’s a minimum)
whose location increases quadratically with $q$ remains visible.
The resonance broadens with $q$, but this is compensated by a deepening
of the resonance peak roughly as $q$ in $\chi_{23}$ and as $q^3$ in $\chi_{33}$.
The resonance in $\chi_{33}$ is caused by the order-parameter
contribution $\chi_{33}-\Pi_{33}$ to the density-density fluctuations (compare the
blue dotted and the blue solid line on the bottom left panel of Fig.~\ref{fig:LU}), in which it is a global
minimum as a function of $\omega$ (rather than a local minimum in  $\chi_{33}$). To emphasize the dispersion
of the resonance, we thus plot on the bottom right panel of Fig.~\ref{fig:LU}, $\Pi_{33}-\chi_{33}$ divided
by $q^3$ in colors as a function of $\omega$ and $q$. The global extremum of $\omega\mapsto\Pi_{33}-\chi_{33}$
is shown as a function of $q$ in white solid line. 
As long as it stays in the window $[2\Delta,\omega_2]$, it varies approximatively quadratically
with $q$.

\begin{figure}[htb]
\includegraphics[width=0.49\textwidth]{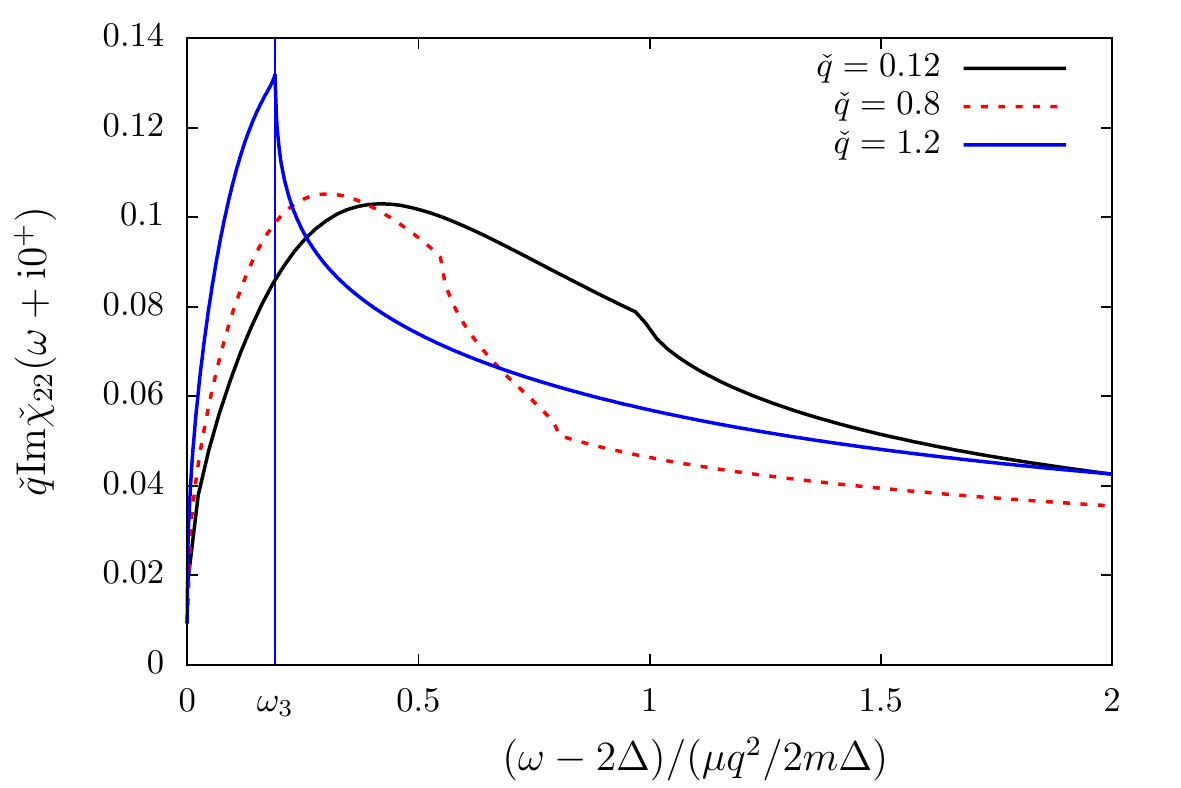}
\includegraphics[width=0.49\textwidth]{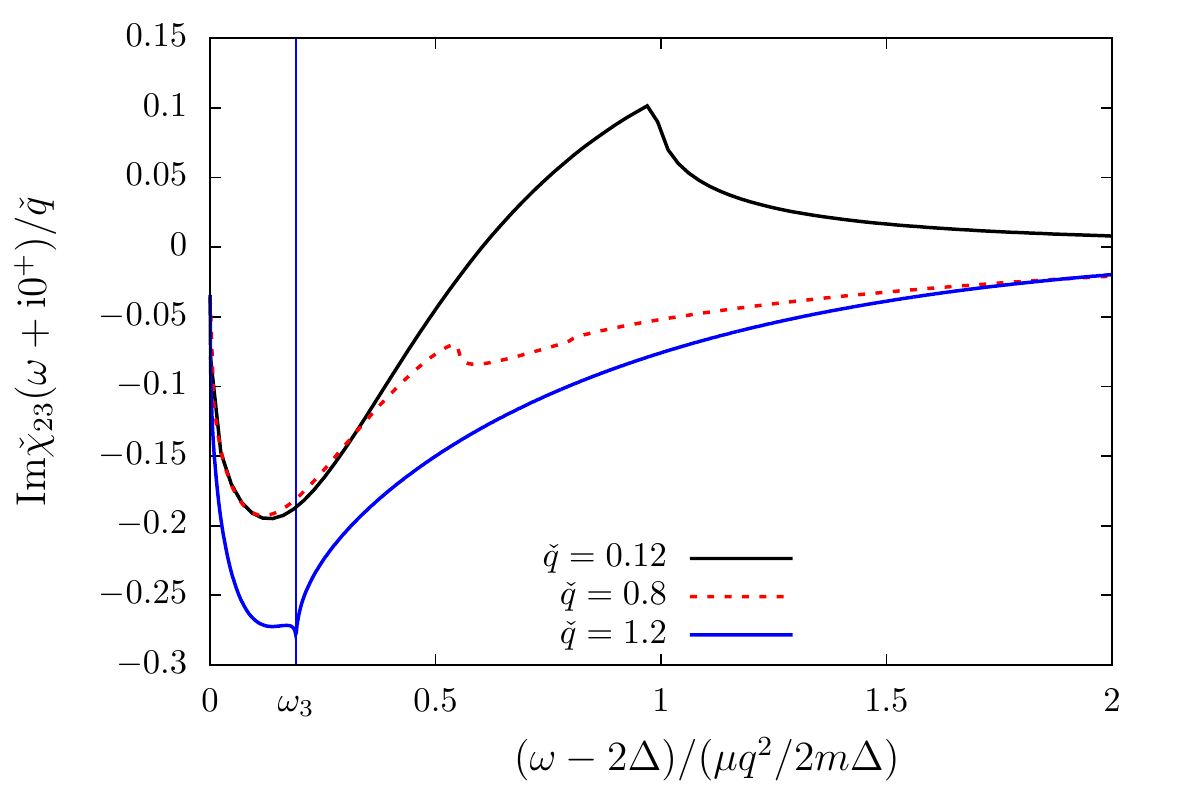}
\includegraphics[width=0.49\textwidth]{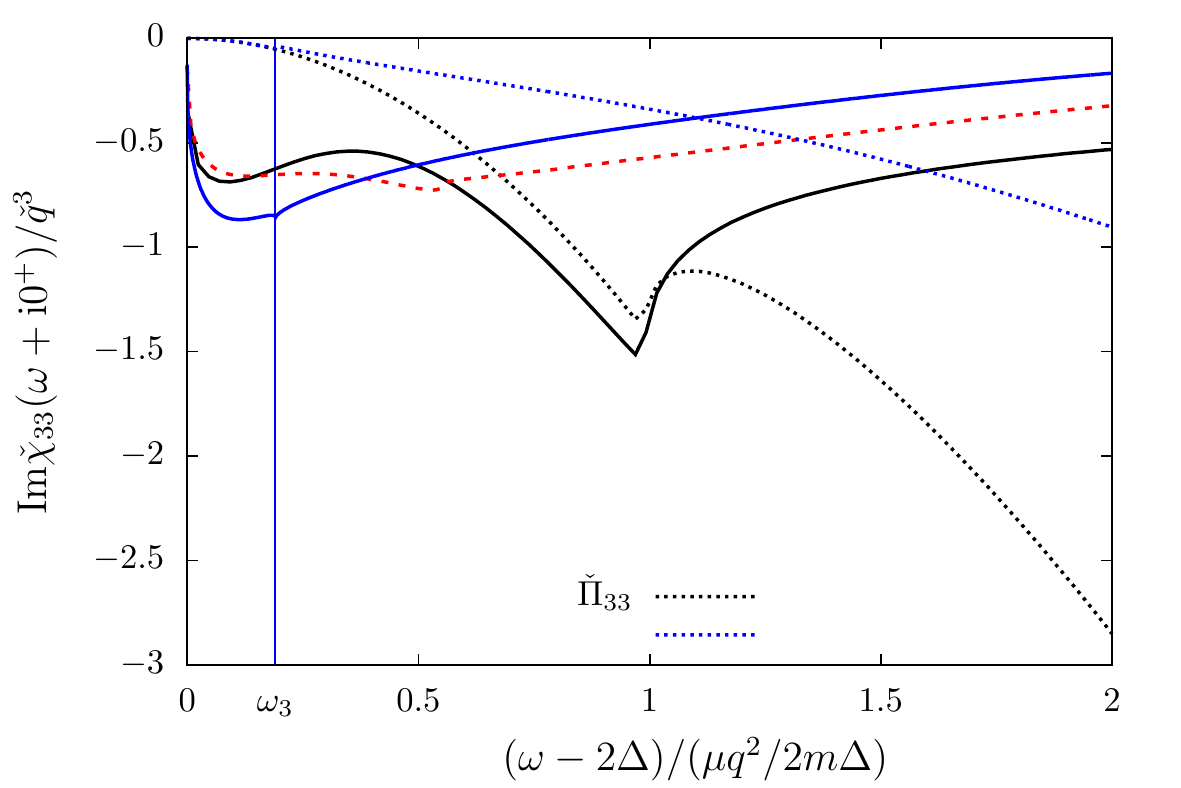}
\includegraphics[width=0.49\textwidth]{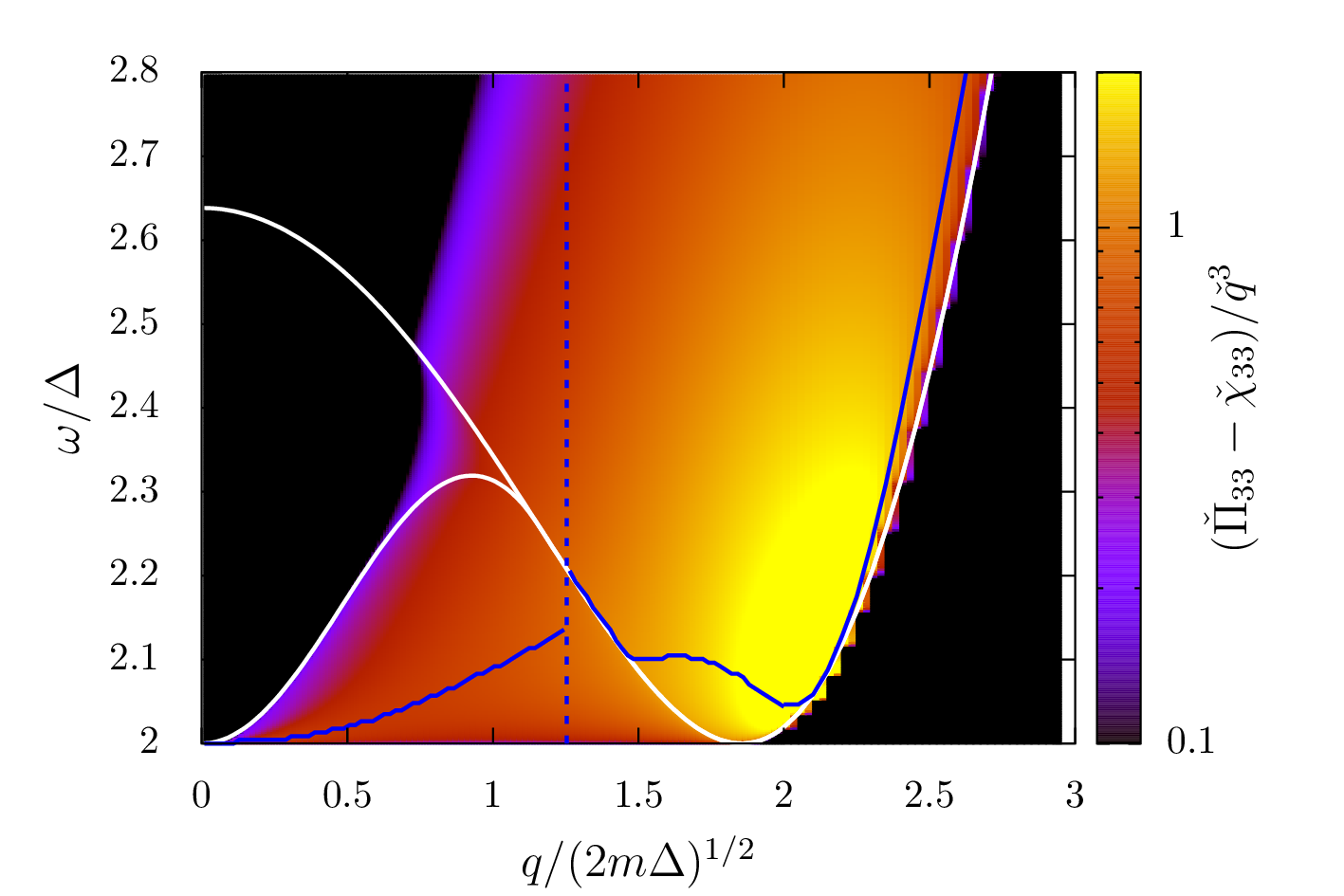}
\caption{\label{fig:LU} The modulus-modulus (top left panel), modulus-density (top right panel) and  density-density (bottom left panel) response functions are shown at unitarity ($\mu/\Delta\simeq0.86$) in function of the reduced drive frequency $(\omega-2\Delta)/(\mu{q}^2/2m\Delta)$ for increasing value of the wavevector $q/\sqrt{2m\Delta}=0.12$ (black solid line), $0.8$ (red dashed line) and $1.2$ (blue solid line). For $q/\sqrt{2m\Delta}=1.2$, we show by a vertical blue line the value of the singularity $\omega_2=\omega_3\simeq 2.23\Delta$ where the shape of the response functions changes dramatically. In the bottom left panel, we also
show the contribution of the pure density-density fluctuations $\Pi_{33}$ to the total density response (black and blue dotted line). 
Bottom right panel: $\check\Pi_{33}-\check\chi_{33}$ is shown 
in colors as a function of $\omega$ and $q$ (the color scale is logarithmic) after division by $\check q^3$. 
The angular points $\omega_2$ and $\omega_3\geq\omega_2$
are superimposed on the color plot as white solid lines. The global extremum of the function $\omega\mapsto\chi_{33}(\omega+i0^+)-\Pi_{33}(\omega+i0^+)$
is shown as a blue solid line. Its location is discontinuous in $\check q\simeq 1.25$ 
(vertical dashed line) after which it coincides with the angular point $\omega_3$ for a range of values of $q$.}
\end{figure}

Contrarily to what happens at weak-coupling, the resonance shape at strong coupling is much distorted
when going through the singularity $\omega_3$. This effect is particularly visible on the modulus-modulus
and modulus-density responses (upper panels of Fig.~\ref{fig:LU}) where the resonance seems broken in $\omega_3$ such that 
the smooth extremum has disappeared in favour of a sharp extremum  in $\omega_3$. On the color plot
of Fig.~\ref{fig:LU}, the quadratic growth of the resonance frequency is also visibly halted
when it encounters the angular point in $\omega_3$.
This is not surprising since the poles found in the analytic continuation through
windows $[2\Delta,\omega_3]$ and $[\omega_3,+\infty[$ are very far apart
in this regime \cite{higgslong}. For the value $q/\sqrt{2m\Delta}=1.2$ used in Fig.~\ref{fig:LU},
the analytic continuation through the interval $[2\Delta,\omega_3]$ has a pole in $z_q/\Delta=1.93-0.41i$.
In the interval  $[\omega_3,+\infty[$, the pole is in $z_q/\Delta=0.86-0.020i$, with a much lower
value of the eigenfrequency and a small damping rate which give this ``upper tail’’ appearance to the response functions
at $\omega>\omega_3$. Above $\omega_3$, the behavior of the response functions 
is in fact similar to what happens in the BEC regime (see below Sec.~\ref{sec:BEC}), with a sharp edge pinned
at $\omega_3$ (which becomes the lower edge of the continuum when $q=2\sqrt{2m\mu}$).

\subsubsection{In the BEC regime}
\label{sec:BEC}

In the BEC regime (that is for us when $\mu<0$), the lower-edge of the pair-breaking continuum
is no longer flat at low $q$, but increases quadratically with $q$. 
Although a pole can be found in the analytic continuation through
the interval $[\omega_3,\infty[$ (the only one available when $\mu<0$),
its real part always stays below $\omega_3$, such that no smooth peak appears
in the response function. Instead there is only a sharp feature
pinned at the lower-edge of the continuum. Fig.~\ref{fig:couleurBEC}, shows
the example of the modulus-modulus response function (the other responses
have a similar behavior) at $\mu/\Delta=-1$ ($1/k_F a\simeq 1.3$).
This sharp feature can hardly be interpreted as a collective mode
and only reflects the incoherent response of the fermionic continuum
when the pairs are tightly bound.

\begin{figure}[htb]
\includegraphics[width=0.6\textwidth]{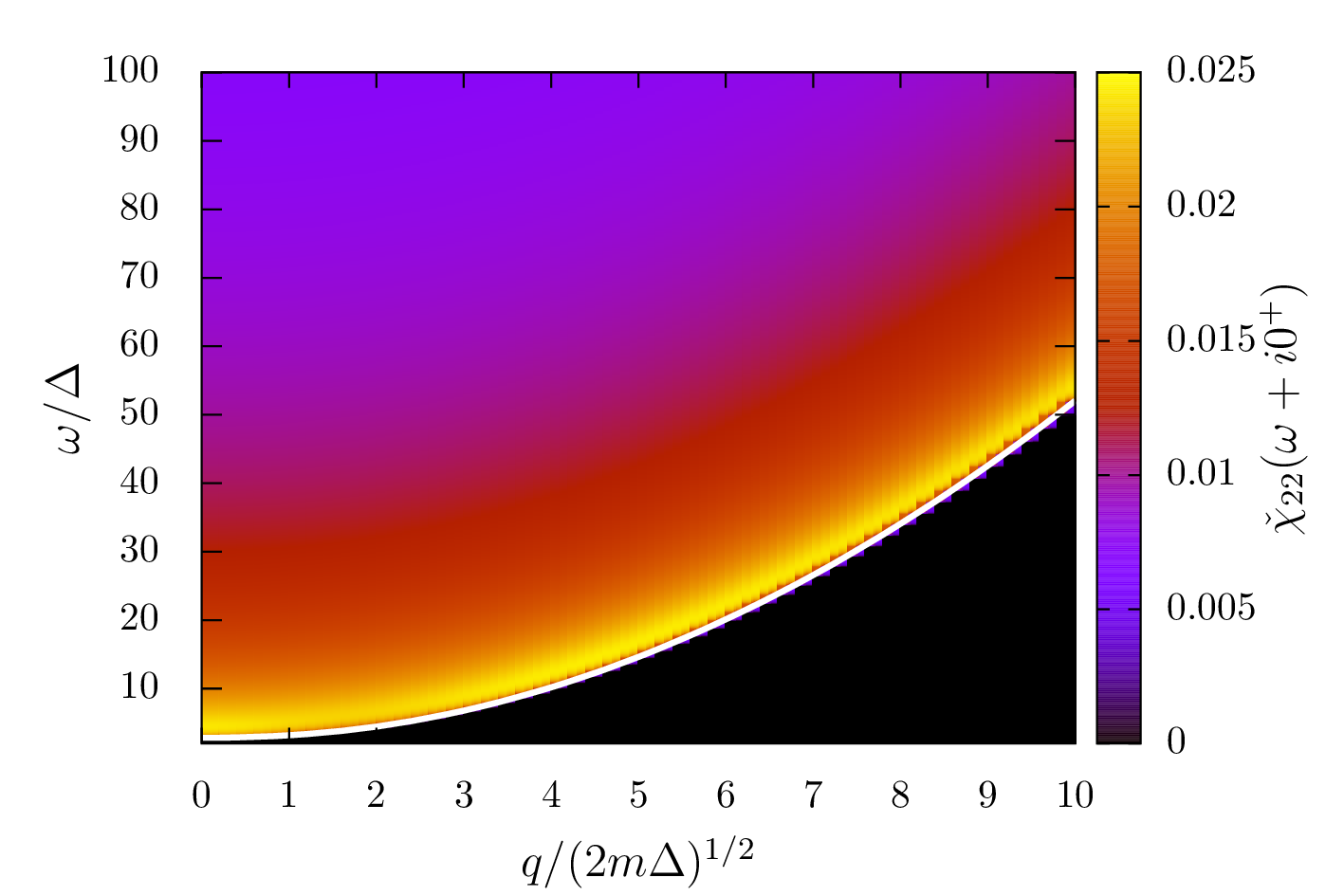}
\caption{\label{fig:couleurBEC} The modulus-modulus response function $\chi_{22}$ in the BEC 
regime ($\mu/\Delta=-1$, $1/k_{\rm F}a\simeq1.3$) is shown in colors 
as a function of the wavevector $q$ and drive frequency $\omega$.
The lower edge $\omega_3=2\sqrt{(\hbar^2q^2/8m+|\mu|)^2+\Delta^2}$
is shown as a white solid line.}
\end{figure}

\subsection{Near $T_c$}

At nonzero temperature and even near $T_c$, we have shown in section \ref{sec:petitsq}
that the Popov-Andrianov resonance exists in the limit $q\to0$ and is almost insensitive
to the quasiparticle-quasihole contributions \eqref{S} to the fluctuation matrix $\Pi$.
This is no longer the case at higher $q$. The angular point $\omega_{\rm ph}$
of the quasiparticle-quasihole continuum in particular destroys the resonance
as it increases (initially linearly) with $q$. This effect
is illustrated on Fig.~\ref{fig:disp} showing the modulus-modulus response
function near $T_c$: at $q\sqrt{2\Delta/m}=0.12$ (orange dashed curve on Fig.~\ref{fig:disp}) 
the lower tail of the resonance is trimmed by the angular point at $\omega_{\rm ph}$,
and at $q\sqrt{2\Delta/m}=0.3$ (long-dashed green curve) it is completely hidden. This can be understood
by a simple reasoning: near $T_c$, $\omega_{\rm ph}$ varies as $q\sqrt{2\mu/m}$ at low $q$ \cite{artlongsk},
such that it reaches $2\Delta$ for $\check q=q/\sqrt{2m\Delta}\approx\sqrt{\Delta/\mu}=O(T_c-T)^{1/4}$. The long
wavelength limit near $T_c$ is thus limited to $q^2/2m\ll \Delta^2/\mu$ (as in the weak-coupling case at $T=0$ see Eq.~(90) in \cite{higgslong}).

\begin{figure}[htb]
\includegraphics[width=0.6\textwidth]{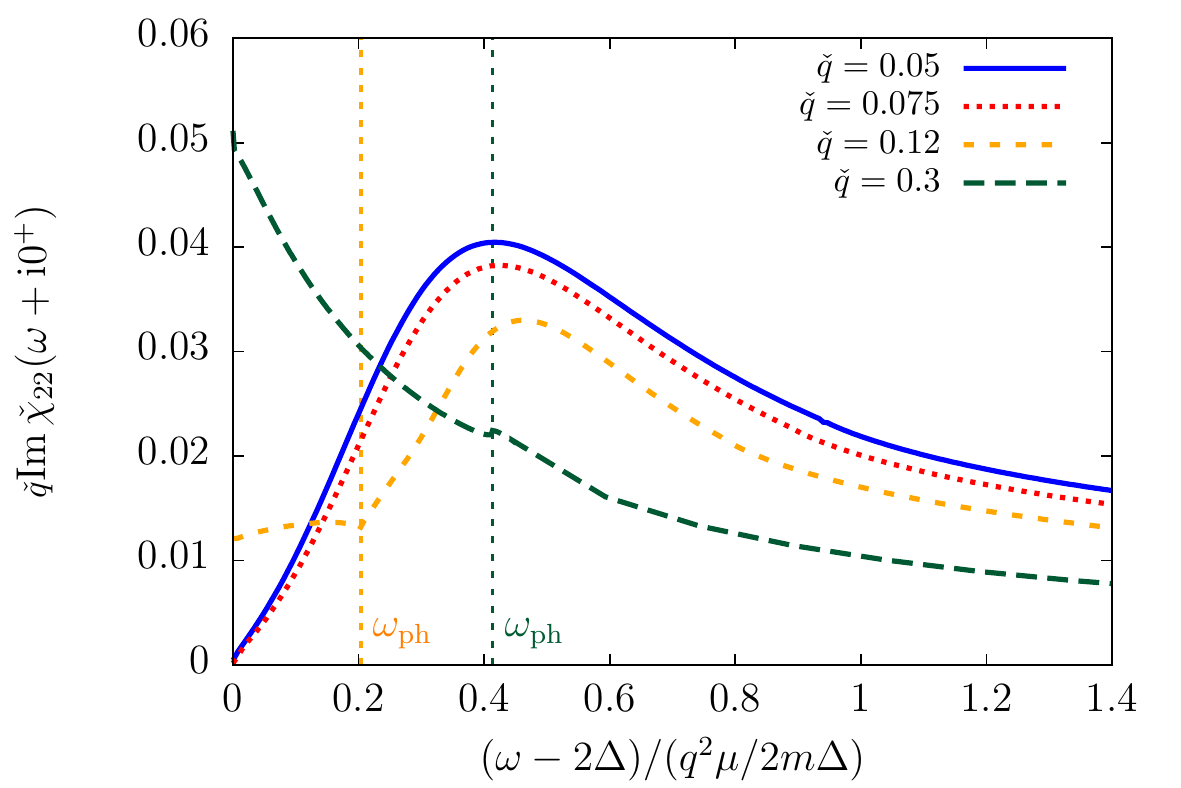}
\caption{\label{fig:disp} The dispersion of the Popov-Andrianov branch in the modulus-modulus response function near the transition temperature ($\Delta/T=0.1$) and in the weak-coupling regime $\mu_c/T_{\rm c}\simeq\mu/T=10$ ($1/k_{\rm F}a\simeq-1.15$ with the mean-field equation-of-state). The modulus-modulus response function is plotted as a function of the reduced drive frequency $(\omega-2\Delta)/({q}^2\mu/2m\Delta)$ for the values of the wavector $q/\sqrt{2m\Delta}= 0.05,\ 0.075,\ 0.12$ and $0.3$ (respectively solid blue, dotted red, dashed orange and long-dashed green lines).
The resonance is well visible at low-$q$ but it disappears as the angular point of the quasiparticle-quasihole continuum $\omega_{\rm ph}$ (shown by a vertical dotted line at $(\omega-2\Delta)/({q}^2\mu/2m\Delta)=0.204$ and $0.413$ for respectively $q/\sqrt{2m\Delta}=0.12$ and $0.3$) rises with $q$.}
\end{figure}

}

\section{Conclusion}

We have computed the response function matrix of a superfluid Fermi gas
in the Random Phase Approximation at nonzero temperature, and used it
to study the observability of the order-parameter collective modes.
We have shown that the appearance of a resonance inside the pair-breaking continuum
associated to the Popov-Andrianov-``Higgs'' mode is a very robust phenomenon which
concerns not only the modulus-modulus response function but also the modulus-density
and density-density responses, which are easier to measure. 
At weak-coupling the resonance is observable at all values of the wavevector $q$
and is only weakly sensitive to the angular points created in the response functions by the changes of structure of the fermionic continuum.
At nonzero temperature, we have shown analytically that
the resonance is not destroyed by the presence of
excited fermionic quasiparticles, and retains approximatively the same shape
as when $T=0$.  It also coexists
with the low-velocity phononic collective mode which RPA predicts near $T_c$. 
The spectral weight of the resonance is enhanced in the
modulus-density and density-density responses when $T$ increases, which should favour its observability.

\begin{acknowledgments}
This research was supported by the Bijzonder Onderzoeksfunds (BOF) of the 
University of Antwerp, the Fonds voor Wetenschappelijk Onderzoek Vlaanderen, 
project G.0429.15.N, and the European Union's Horizon 2020 research and innovation program under the Marie Sk\l odowska-Curie grant agreement number 665501.
\end{acknowledgments}

\appendix

\section{Derivation of the equations of motion}
\label{app:eom}

We give here a few additional steps leading to the equations
of motion (\ref{systRPA1}--\ref{systRPA4}).
In the particle basis, 
the equations of motion take the form:
\bea
\!\!\!\!\!\!\!\!\!\!\!\!\deriv{\kq{d}}   &\!\!=& \!\!\! \xi^+_{\kk\qq} \kq{d} -\Delta_0 (\hat{n}_{\kk }^{\qq}+\hat{\bar{n}}_{\kk}^{\qq}) + z_{\kk\qq}^+  (\hat{\Delta}^\qq+\phi(\qq))
+  d_{\kk\qq}^+ \frac{g_0\hat{n}^{\qq}_\uparrow+g_0\hat{n}^{\qq}_\downarrow+u_+(\qq)}{2}  -  d_{\kk\qq}^-  \frac{g_0\hat{n}^{\qq}_\uparrow-g_0\hat{n}^{\qq}_\downarrow  -u_-(\qq) }{2} \label{systRPAparticules1}\\
\!\!\!\!\!\!\!\!\!\!\!\!\deriv{\kqb{d}}  &\!\!=& \!\!\! -\xi^+_{\kk\qq} \kqb{d} +\Delta_0 (\hat{n}_{\kk }^{\qq}+\hat{\bar{n}}_{\kk}^{\qq}) - z_{\kk\qq}^+ (\hat{\bar{\Delta}}^\qq+\bar\phi(\qq))
-  d_{\kk\qq}^+ \frac{g_0\hat{n}^{\qq}_\uparrow+g_0\hat{n}^{\qq}_\downarrow+u_+(\qq)}{2} -  d_{\kk\qq}^-  \frac{g_0\hat{n}^{\qq}_\uparrow-g_0\hat{n}^{\qq}_\downarrow  -u_-(\qq) }{2}  \label{systRPAparticules2} \\
\!\!\!\!\!\!\!\!\!\!\!\!\deriv{\hat{n}_{\kk}^{\qq}} &\!\!=& \!\!\! - \xi^-_{\kk\qq} \hat{n}_{\kk }^{\qq} -\Delta_0 (\hat{d}_{\kk}^{\qq}-\hat{\bar{d}}_{\kk}^{\qq}) -  z_{\kk\qq}^- (g_0\hat{n}^{\qq}_\downarrow+u_\uparrow(\qq) )
 + d_{\kk\qq}^-  \frac{\hat{\Delta}^\qq+\hat{\bar{\Delta}}^\qq+\phi_+(\qq)}{2}+  d_{\kk\qq}^+  \frac{\hat{\Delta}^\qq-\hat{\bar{\Delta}}^\qq+\phi_-(\qq)}{2} \label{systRPAparticules3} \\
\!\!\!\!\!\!\!\!\!\!\!\!\deriv{\hat{\bar{n}}_{\kk }^{\qq}} &\!\!=& \!\!\! \xi^-_{\kk\qq} \hat{\bar{n}}_{\kk }^{\qq} -\Delta_0 (\hat{d}_{\kk}^{\qq}-\hat{\bar{d}}_{\kk}^{\qq}) +  z_{\kk \qq }^- (g_0\hat{n}^{\qq}_\uparrow +u_\downarrow(\qq))\label{systRPAparticules4}
 - d_{\kk\qq}^-  \frac{\hat{\Delta}^\qq+\hat{\bar{\Delta}}^\qq+\phi_+(\qq)}{2}+  d_{\kk\qq}^+  \frac{\hat{\Delta}^\qq-\hat{\bar{\Delta}}^\qq+\phi_-(\qq)}{2},
\eea
where we generalize the notations of Refs.~\cite{Anderson1958,TheseHK} to nonzero temperature:
\bea
z_{\kk\qq}^\pm (T)       &=& \bb{\frac{1}{2}-\meanv{\hat n_{\kk_+}^\zero}_T} \pm \bb{\frac{1}{2}-\meanv{\hat n_{\kk_-}^\zero}_T} = \frac{U_{\kk_+}^2-V_{\kk_+}^2}{2}(1-2f_{\kk_+})\pm \frac{U_{\kk_-}^2-V_{\kk_-}^2}{2}(1-2f_{\kk_-}) \notag \\ 
&=& \bbcro{ \frac{\xi_{\kk_+}}{2 \epsilon_{\kk_+}} \pm \frac{\xi_{\kk_-}}{2\epsilon_{\kk_-}}} (1-f_{\kk_+}-f_{\kk_-}) -  \bbcro{ \frac{\xi_{\kk_+}}{2 \epsilon_{\kk_+}} \mp \frac{\xi_{\kk_-}}{2\epsilon_{\kk_-}}} (f_{\kk_+}-f_{\kk_-})  \label{zkq}  \\
d_{\kk\qq}^\pm(T) &=&\meanv{\hat{d}_{\kk+\qq/2}^\zero}_T\pm\meanv{\hat{d}_{\kk-\qq/2}^\zero}_T = - \bbcro{U_{\kk_+} V_{\kk_+}(1-2f_{\kk_+}) \pm U_{\kk_-} V_{\kk_-}(1-2f_{\kk_-})} \notag \\ 
&=& - \bbcro{ \frac{\Delta}{2 \epsilon_{\kk_+}} \pm \frac{\Delta}{2\epsilon_{\kk_-}}} (1-f_{\kk_+}-f_{\kk_-}) +  \bbcro{ \frac{\Delta}{2 \epsilon_{\kk_+}} \mp \frac{\Delta}{2\epsilon_{\kk_-}}} (f_{\kk_+}-f_{\kk_-}).\label{dkq} 
\eea
Adding and subtracting Eq.~\eqref{systRPAparticules1} to \eqref{systRPAparticules2} and Eq.~\eqref{systRPAparticules3} to \eqref{systRPAparticules4} and performing the change of basis \eqref{passage_qpart} (one can use the explicit relations given in Appendix C of \cite{TheseHK}) yields the equations
of motion (\ref{systRPA1}--\ref{systRPA4}) in the quasiparticle basis.
Rederiving with respect to time yields:
\bea
-(\epsilon_{\kk\qq}^+)^2 \kq{y}-\frac{\dd^2 \kq{y}}{\dd t^2} &=&  (1-f_{\kk_+}-f_{\kk_-}) \bbcrol{W_{\kk\qq}^- \ii\hbar\partial_t \bb{\delta\hat{\Delta}^\qq+\delta\hat{\bar{\Delta}}^\qq + \phi_+(\qq)} -  w_{\kk\qq}^+ \ii\hbar\partial_t \bb{g_0\bbcro{\delta\hat{n}_\uparrow^\qq+\delta\hat{{n}}_\downarrow^\qq}+u_+(\qq)} } \notag\\
&& \bbcror{+ \epsilon_{\kk\qq}^+ W_{\kk\qq}^+  \bb{\hat{\Delta}^\qq-\hat{\bar{\Delta}}^\qq+\phi_-(\qq)} -  \epsilon_{\kk\qq}^+ w_{\kk\qq}^-  \bb{g_0\bbcro{\hat{n}_\uparrow^\qq-\hat{{n}}_\downarrow^\qq}-u_-(\qq)}} \\
-(\epsilon_{\kk\qq}^-)^2 \kq{h} - \frac{\dd^2 \kq{h}}{\dd t^2} &=& (f_{\kk_+}-f_{\kk_-}) \bbcrol{w_{\kk\qq}^+ \ii\hbar\partial_t \bb{\delta\hat{\Delta}^\qq+\delta\hat{\bar{\Delta}}^\qq + \phi_+(\qq)} +  W_{\kk\qq}^- \ii\hbar\partial_t \bb{g_0\bbcro{\delta\hat{n}_\uparrow^\qq+\delta\hat{{n}}_\downarrow^\qq}+u_+(\qq)} } \notag\\ 
&& \bbcror{+ \epsilon_{\kk\qq}^- w_{\kk\qq}^-  \bb{\hat{\Delta}^\qq-\hat{\bar{\Delta}}^\qq+\phi_-(\qq)} +  \epsilon_{\kk\qq}^- W_{\kk\qq}^+  \bb{g_0\bbcro{\hat{n}_\uparrow^\qq-\hat{{n}}_\downarrow^\qq}-u_-(\qq)}} \\
-(\epsilon_{\kk\qq}^+)^2 \kq{s} - \frac{\dd^2 \kq{s}}{\dd t^2} &=&  (1-f_{\kk_+}-f_{\kk_-}) \bbcrol{W_{\kk\qq}^+  \ii\hbar\partial_t \bb{\hat{\Delta}^\qq-\hat{\bar{\Delta}}^\qq+\phi_-(\qq)} -  w_{\kk\qq}^- \ii\hbar\partial_t \bb{g_0\bbcro{\hat{n}_\uparrow^\qq-\hat{{n}}_\downarrow^\qq}-u_-(\qq)} } \notag\\
&& \bbcror{+ \epsilon_{\kk\qq}^+ W_{\kk\qq}^-  \bb{\delta\hat{\Delta}^\qq+\delta\hat{\bar{\Delta}}^\qq + \phi_+(\qq)} -  \epsilon_{\kk\qq}^+ w_{\kk\qq}^+   \bb{g_0\bbcro{\delta\hat{n}_\uparrow^\qq+\delta\hat{{n}}_\downarrow^\qq}+u_+(\qq)} } \\
-(\epsilon_{\kk\qq}^-)^2 \kq{m}- \frac{\dd^2 \kq{m}}{\dd t^2} &=&  - (f_{\kk_+}-f_{\kk_-}) \bbcrol{w_{\kk\qq}^- \ii\hbar\partial_t \bb{\hat{\Delta}^\qq-\hat{\bar{\Delta}}^\qq+\phi_-(\qq)} +  W_{\kk\qq}^+ \ii\hbar\partial_t  \bb{g_0\bbcro{\hat{n}_\uparrow^\qq-\hat{{n}}_\downarrow^\qq}-u_-(\qq)} } \notag\\ 
&& \bbcror{+ \epsilon_{\kk\qq}^- w_{\kk\qq}^+  \bb{\delta\hat{\Delta}^\qq+\delta\hat{\bar{\Delta}}^\qq + \phi_+(\qq)}  +  \epsilon_{\kk\qq}^- W_{\kk\qq}^- \bb{g_0\bbcro{\delta\hat{n}_\uparrow^\qq+\delta\hat{{n}}_\downarrow^\qq}+u_+(\qq)} }.
\eea
We resum this system to form the collective quantities (\ref{eq:op_col_1}--\ref{eq:op_col_4})
and derive the $4\times4$ linear system \eqref{4par4}.

\section{Numerical calculation of the response functions}
\label{app:numerique}
\res{
To numerically compute the fluctuation matrix $\Pi$, we first compute its spectral
density:
\be
\mbox{Im}\check\Pi_{ij}(\omega+i0^+)=-\pi\bb{\rho_{ij}^{\rm (pp)}(\omega)+\rho_{ij}^{\rm (ph)}(\omega)},
\ee
where $\rho_{ij}^{\rm (pp)}$ and $\rho_{ij}^{\rm (ph)}(\omega)$ are respectively the contributions of 
the quasiparticle-quasiparticle integral $\Sigma$ and quasparticle-quasihole integral $S$ to the spectral density 
of $\Pi_{ij}$. Denoting $u=\kk\cdot\qq/kq$, and restricting, without loss of generality, to $\omega>0$, we have, explicitly:
\be
\rho_{ij}^{\rm (pp)}(\omega)=\frac{2\pi\Delta}{k_\Delta^3}\eta_{ij}\int_0^{+\infty} k^2 dk \int_0^1 du a_{i,\kk\qq} a_{j,\kk\qq} (1-f_{\kk+\qq/2}-f_{\kk-\qq/2}) \delta(\omega-\epsilon_{\kk\qq}^+),
\label{rhoijpp}
\ee
We have introduced $k_\Delta=\sqrt{2m\Delta}$, the coefficients $a_{1,\kk\qq}=W_{\kk,\qq}^+$, $a_{2,\kk\qq}=W_{\kk,\qq}^-$ and $a_{3,\kk\qq}=w_{\kk,\qq}^+$ and the signs $\eta_{ij}$, read from \eqref{Pi}
\be
\eta=\begin{pmatrix} 1&1&-1 \\ 1&1&-1\\ -1&-1&1\end{pmatrix}.
\ee 
For the particle-hole contribution, we have
\be
\rho_{ij}^{\rm (ph)}(\omega)=-\frac{2\pi\Delta}{k_\Delta^3}\int_0^{+\infty} k^2 dk \int_0^1 du b_{i,\kk\qq} b_{j,\kk\qq} (f_{\kk+\qq/2}-f_{\kk-\qq/2}) \bbcro{\delta(\omega-\epsilon_{\kk\qq}^-) - \sigma_{ij}\delta(\omega+\epsilon_{\kk\qq}^-) }.
\label{rhoijph}
\ee
Here, $b_{1,\kk\qq}=w_{\kk,\qq}^-$, $b_{2,\kk\qq}=w_{\kk,\qq}^+$ and $b_{3,\kk\qq}=W_{\kk,\qq}^-$ and the sign $\sigma_{ij}$ is $+1$ for
$S^\epsilon$ matrix elements and $-1$ for the $S^\omega$:
\be
\sigma=\begin{pmatrix} 1&-1&-1 \\ -1&1&1\\ -1&1&1\end{pmatrix}.
\ee 
In \eqref{rhoijpp} and \eqref{rhoijph}, we have used the symmetry or antisymmetry of the coefficients 
$a_{i,\kk\qq} a_{j,\kk\qq}$ and $b_{i,\kk\qq} b_{j,\kk\qq}$ with respect to the
exchange $u\leftrightarrow-u$ to restrict the integral to $u>0$.

In the quasiparticle-quasiparticle spectral density \eqref{rhoijpp}, we give the resonance angle:
\be
u_r= \frac{m\omega}{kq} \bb{\frac{\xi^2-(\omega^2-4\Delta^2)/4}{\xi^2-\omega^2/4}}^{1/2} \qquad \mbox{with} \qquad \xi=k^2/2m+q^2/8m-\mu
\label{ur}
\ee
For $\omega_1<\omega<\omega_2$, this quantity is comprised between $[0,1]$ (such that the resonance in \eqref{rhoijpp} is reached) for $k\in[k_1,k_2]$ with $k_1$ and $k_2$ solutions of $(\epsilon_{\kk+\qq/2}+\epsilon_{\kk-\qq/2})\vert_{u=0}=\omega$. For $\omega_2<\omega<\omega_3$ the resonance is reached for  $k\in[k_1,k_1’]$ and $k\in[k_2’,k_2]$ with $k_1’$ and $k_2’$ solutions of $\epsilon_{k+q/2}+\epsilon_{k-q/2}=\omega$. Finally for $\omega>\omega_3$ the resonance is reached for $k\in[k_2’,k_2]$ only.
Using the variable $y=2\xi/\omega$ instead of the wavenumber $k$, and $t=\mbox{argch}(\omega/2\Delta)$ instead of the drive frequency, then using the Dirac
delta to integrate analytically over the scattering angle $u$, we have:
\be
\rho_{ij}^{\rm (pp)}(\omega) =\frac{\pi}{4\check q \ch t} \begin{cases}  \int_{-\textrm{th} t}^{\textrm{th} t}dy \frac{\tilde W_{ij}(y)(1-\tilde f_+(y)-\tilde f_-(y))}{\sqrt{(\textrm{th}^2 t-y^2)(1-y^2)}} \qquad \mbox{if} \qquad \omega_1<\omega<\omega_2 \\
\bbcro{\int_{-\textrm{th} t}^{y_1’}dy+\int_{y_2’}^{\textrm{th} t}dy} \frac{\tilde W_{ij}(y)(1-\tilde f_+(y)-\tilde f_-(y))}{\sqrt{(\textrm{th}^2 t-y^2)(1-y^2)}} \qquad \mbox{if} \qquad \omega_2<\omega<\omega_3 \\
\int_{y_2’}^{\textrm{th} t} \frac{\tilde W_{ij}(y)(1-\tilde f_+(y)-\tilde f_-(y))}{\sqrt{(\textrm{th}^2 t-y^2)(1-y^2)}} \qquad \mbox{if} \qquad \omega>\omega_3 \end{cases}
\ee
where $y_1’$ and $y_2’$ are deduced from $k_1’$ and $k_2’$ by the change of variable given above, and the functions $W_{ij}$ are:
\bea
\tilde W_{11}(y)&=&\frac{2}{1-y^2}\\
\tilde W_{22}(y)&=&\frac{2y^2}{1-y^2}\\
\tilde W_{12}(y)&=&\frac{2y}{1-y^2}\\
\tilde W_{13}(y)&=&{2\ch t}\\
\tilde W_{23}(y)&=&{2y\ch t}\\
\tilde W_{33}(y)&=&2\ch^2 t ({1-y^2}).
\eea
In our integration variables, the Fermi-Dirac occupation numbers have the expression
\bea
\tilde f_\pm(y)=1/\bb{1+\exp \bbcro{\sqrt{\ch^2 t (y\pm kqu_r/m\omega)^2+1}\times\Delta/T}}.
\eea

In the quasiparticle-quasihole spectral density \eqref{rhoijph}, we give the resonance angle expressed in terms of 
$\xi=k^2/2m+q^2/8m-\mu$ has the expression \eqref{ur} given above.
Whatever the value of $\omega$ this angle exists (\textit{i.e} $u_r\in[0,1]$) for $k\in[\tilde k_1,\infty[$, with $\tilde k_1$ the solution of $\epsilon_{k+q/2}-\epsilon_{k-q/2}=\omega$. When $\omega<\omega_{\rm ph}$, it also exists for $k\in[\tilde k_3,\tilde k_2]$, with $\tilde k_3,\tilde k_2$ the two solutions of $\epsilon_{k+q/2}-\epsilon_{k-q/2}=-\omega$.
Using the variable $y=\omega/2\xi$  instead of the wavenumber $k$, and $t=\mbox{arccos}(\omega/2\Delta)$ instead of the drive frequency, then using the Dirac
delta to integrate analytically over the scattering angle $u$, we have:
\be
\rho_{ij}^{\rm (ph)}(\omega) = -\frac{\pi}{4\tilde q \cos t} \bbcro{\int_0^{\tilde y_1}dy-\Theta(\omega_{\rm ph}-\omega)\int_{\tilde y_3}^{\tilde y_2}dy} \frac{\tilde w_{ij}(y)(\tilde f_+(y)-\tilde f_-(y))}{\sqrt{(1+y^2\tan^2 t)(1-y^2)}},
\ee
where $\tilde y_i$ is related to $\tilde k_i$ by the change of variable given above, and the functions $w_{ij}$ are
\bea
\tilde w_{11}(y)&=&\frac{2y^2}{1-y^2}\\
\tilde w_{22}(y)&=&\frac{2}{1-y^2}\\
\tilde w_{12}(y)&=&\frac{2y}{1-y^2}\\
\tilde w_{13}(y)&=&{2\cos t}\\
\tilde w_{23}(y)&=&\frac{2\cos t}{y}\\
\tilde w_{33}(y)&=&2\cos^2 t \frac{1-y^2}{y^2}.
\eea
Here, the Fermi-Dirac occupation numbers  have the expression
\bea
\tilde f_\pm(y)=1/(1+\exp \bbcro{\sqrt{\cos^2 t (1/y\pm kqu_r/m\omega)^2+1}\times\Delta/T}).
\eea

Finally, to compute the full function, we use the spectral density to integrate over energies:
\be
\check\Pi_{ij}(\omega_0)=\int_0^{+\infty} d\omega \bbcro{\frac{\rho_{ij}(\omega)}{\omega_0-\omega}-\sigma_{ij}\frac{\rho_{ij}(\omega)}{\omega_0+\omega}+(\delta_{i1}\delta_{j1}+\delta_{i2}\delta_{j2})\frac{\pi\sqrt{\omega}}{\sqrt{8\bb{1+\bb{\omega/2-\mu/\Delta}^2}}}}.
\ee
In $\check\Pi_{11}$ and $\check\Pi_{22}$, the divergence at large $\omega$ is regularized by the counter-term $4\pi\int_0^{\infty}\frac{k^2dk}{k_\Delta^3}\frac{\Delta}{2\epsilon_k}=\int_0^{+\infty} d\omega \frac{\pi\sqrt{\omega}}{8\bb{1+\bb{\omega/2-\mu/\Delta}^2}}$.}
\bibliography{/Users/hkurkjian/Documents/biblio}

\providecommand*\hyphen{-}
\begin{thebibliography}{52}
\providecommand{\natexlab}[1]{#1}
\providecommand{\url}[1]{\texttt{#1}}
\expandafter\ifx\csname urlstyle\endcsname\relax
  \providecommand{\doi}[1]{doi: #1}\else
  \providecommand{\doi}{doi: \begingroup \urlstyle{rm}\Url}\fi

\bibitem[{Fetter} and {Walecka}(1971)]{FetterWalecka}
Alexander~L. {Fetter} and John~Dirk {Walecka}.
\newblock \emph{{Quantum theory of many-particle systems}}.
\newblock McGraw-Hill, San Francisco, 1971.

\bibitem[Nozières(1963)]{Nozieres1963}
Philippe Nozières.
\newblock \emph{Le problème à $N$ corps\,:\,propriétés générales des gaz
  de fermions}.
\newblock Dunod, Paris, 1963.

\bibitem[Cohen-Tannoudji et~al.(1988)Cohen-Tannoudji, Dupont-Roc, and
  Grynberg]{CohenChapIIIC}
C.~Cohen-Tannoudji, J.~Dupont-Roc, and G.~Grynberg.
\newblock \emph{{Processus d'interaction entre photons et atomes}}, chapter
  III. \'Etude non perturbative des amplitudes de transition.
\newblock InterEditions et \'Editions du {CNRS}, Paris, 1988.

\bibitem[Greiner et~al.(2003)Greiner, Regal, and Jin]{Jin2003}
Markus Greiner, Cindy~A. Regal, and Deborah~S. Jin.
\newblock {Emergence of a molecular Bose-Einstein condensate from a Fermi gas}.
\newblock \emph{Nature}, 426\penalty0 (6966):\penalty0 537--540, December 2003.
\newblock URL \url{http://dx.doi.org/10.1038/nature02199}.

\bibitem[Zwierlein et~al.(2003)Zwierlein, Stan, Schunck, Raupach, Gupta,
  Hadzibabic, and Ketterle]{Ketterle2003}
M.~W. Zwierlein, C.~A. Stan, C.~H. Schunck, S.~M.~F. Raupach, S.~Gupta,
  Z.~Hadzibabic, and W.~Ketterle.
\newblock {Observation of Bose-Einstein Condensation of Molecules}.
\newblock \emph{Phys. Rev. Lett.}, 91:\penalty0 250401, December 2003.
\newblock \doi{10.1103/PhysRevLett.91.250401}.
\newblock URL \url{http://link.aps.org/doi/10.1103/PhysRevLett.91.250401}.

\bibitem[Jochim et~al.(2003)Jochim, Bartenstein, Altmeyer, Hendl, Riedl, Chin,
  Hecker~Denschlag, and Grimm]{Grimm2003}
S.~Jochim, M.~Bartenstein, A.~Altmeyer, G.~Hendl, S.~Riedl, C.~Chin,
  J.~Hecker~Denschlag, and R.~Grimm.
\newblock {Bose-Einstein Condensation of Molecules}.
\newblock \emph{Science}, 302\penalty0 (5653):\penalty0 2101--2103, 2003.
\newblock \doi{10.1126/science.1093280}.
\newblock URL \url{http://www.sciencemag.org/content/302/5653/2101.abstract}.

\bibitem[Zwierlein et~al.(2005)Zwierlein, Abo-Shaeer, Schirotzek, Schunck, and
  Ketterle]{Ketterle2005}
M.~W. Zwierlein, J.~R. Abo-Shaeer, A.~Schirotzek, C.~H. Schunck, and
  W.~Ketterle.
\newblock {Vortices and superfluidity in a strongly interacting Fermi gas}.
\newblock \emph{Nature}, 435\penalty0 (7045):\penalty0 1047--1051, June 2005.

\bibitem[Joseph et~al.(2007)Joseph, Clancy, Luo, Kinast, Turlapov, and
  Thomas]{Thomas2007}
J.~Joseph, B.~Clancy, L.~Luo, J.~Kinast, A.~Turlapov, and J.~E. Thomas.
\newblock {Measurement of Sound Velocity in a Fermi Gas near a Feshbach
  Resonance}.
\newblock \emph{Phys. Rev. Lett.}, 98:\penalty0 170401, April 2007.
\newblock \doi{10.1103/PhysRevLett.98.170401}.
\newblock URL \url{https://link.aps.org/doi/10.1103/PhysRevLett.98.170401}.

\bibitem[Schirotzek et~al.(2008)Schirotzek, Shin, Schunck, and
  Ketterle]{Ketterle2008}
Andr\'e Schirotzek, Yong-il Shin, Christian~H. Schunck, and Wolfgang Ketterle.
\newblock {Determination of the Superfluid Gap in Atomic Fermi Gases by
  Quasiparticle Spectroscopy}.
\newblock \emph{Phys. Rev. Lett.}, 101:\penalty0 140403, October 2008.
\newblock \doi{10.1103/PhysRevLett.101.140403}.
\newblock URL \url{http://link.aps.org/doi/10.1103/PhysRevLett.101.140403}.

\bibitem[Nascimb{\`e}ne et~al.(2010)Nascimb{\`e}ne, Navon, Jiang, Chevy, and
  Salomon]{Salomon2010}
S.~Nascimb{\`e}ne, N.~Navon, K.~J. Jiang, F.~Chevy, and C.~Salomon.
\newblock {Exploring the thermodynamics of a universal Fermi gas}.
\newblock \emph{Nature}, 463\penalty0 (7284):\penalty0 1057--1060, February
  2010.
\newblock URL \url{http://dx.doi.org/10.1038/nature08814}.

\bibitem[Ku et~al.(2012)Ku, Sommer, Cheuk, and Zwierlein]{Zwierlein2012}
Mark J.~H. Ku, Ariel~T. Sommer, Lawrence~W. Cheuk, and Martin~W. Zwierlein.
\newblock {Revealing the Superfluid Lambda Transition in the Universal
  Thermodynamics of a Unitary Fermi Gas}.
\newblock \emph{Science}, 335\penalty0 (6068):\penalty0 563--567, 2012.
\newblock \doi{10.1126/science.1214987}.
\newblock URL \url{http://www.sciencemag.org/content/335/6068/563.abstract}.

\bibitem[Sidorenkov et~al.(2013)Sidorenkov, Tey, Grimm, Hou, Pitaevskii, and
  Stringari]{Stringari2013}
Leonid~A. Sidorenkov, Meng~Khoon Tey, Rudolf Grimm, Yan-Hua Hou, Lev
  Pitaevskii, and Sandro Stringari.
\newblock {Second sound and the superfluid fraction in a {F}ermi gas with
  resonant interactions}.
\newblock \emph{Nature}, 498\penalty0 (7452):\penalty0 78--81, June 2013.

\bibitem[Hoinka et~al.(2017)Hoinka, Dyke, Lingham, Kinnunen, Bruun, and
  Vale]{Vale2017}
Sascha Hoinka, Paul Dyke, Marcus~G. Lingham, Jami~J. Kinnunen, Georg~M. Bruun,
  and Chris~J. Vale.
\newblock {Goldstone mode and pair-breaking excitations in atomic Fermi
  superfluids}.
\newblock \emph{Nature Physics}, 13:\penalty0 943--946, June 2017.
\newblock URL \url{http://dx.doi.org/10.1038/nphys4187}.

\bibitem[Bardeen et~al.(1957)Bardeen, Cooper, and Schrieffer]{BCS1957}
J.~Bardeen, L.~N. Cooper, and J.~R. Schrieffer.
\newblock {Theory of Superconductivity}.
\newblock \emph{Phys. Rev.}, 108:\penalty0 1175--1204, December 1957.
\newblock \doi{10.1103/PhysRev.108.1175}.
\newblock URL \url{http://link.aps.org/doi/10.1103/PhysRev.108.1175}.

\bibitem[Haussmann(1993)]{Haussmann1993}
R.~Haussmann.
\newblock {Crossover from BCS superconductivity to Bose-Einstein condensation:
  A self-consistent theory}.
\newblock \emph{Zeitschrift f{\"u}r Physik B Condensed Matter}, 91\penalty0
  (3):\penalty0 291--308, Sep 1993.
\newblock ISSN 1431-584X.
\newblock \doi{10.1007/BF01344058}.
\newblock URL \url{https://doi.org/10.1007/BF01344058}.

\bibitem[Haussmann et~al.(2009)Haussmann, Punk, and Zwerger]{Zwerger2009}
Rudolf Haussmann, Matthias Punk, and Wilhelm Zwerger.
\newblock Spectral functions and rf response of ultracold fermionic atoms.
\newblock \emph{Physical Review A}, 80\penalty0 (6):\penalty0 063612, 2009.
\newblock \doi{10.1103/PhysRevA.80.063612}.

\bibitem[Van~Loon et~al.(2020)Van~Loon, Tempere, and Kurkjian]{sennepol}
Senne Van~Loon, Jacques Tempere, and Hadrien Kurkjian.
\newblock Beyond mean-field corrections to the quasiparticle spectrum of
  superfluid fermi gases.
\newblock \emph{Phys. Rev. Lett.}, 124:\penalty0 073404, Feb 2020.
\newblock \doi{10.1103/PhysRevLett.124.073404}.
\newblock URL \url{https://link.aps.org/doi/10.1103/PhysRevLett.124.073404}.

\bibitem[Schmid and Sch\"on(1975)]{Schon1975}
Albert Schmid and Gerd Sch\"on.
\newblock {Collective Oscillations in a Dirty Superconductor}.
\newblock \emph{Phys. Rev. Lett.}, 34:\penalty0 941--943, April 1975.
\newblock \doi{10.1103/PhysRevLett.34.941}.
\newblock URL \url{https://link.aps.org/doi/10.1103/PhysRevLett.34.941}.

\bibitem[Andrianov and Popov(1976)]{Popov1976}
V.~A. Andrianov and V.~N. Popov.
\newblock {Gidrodinamičeskoe dejstvie i Boze-spektr sverhtekučih
  Fermi-sistem}.
\newblock \emph{Teoreticheskaya i Matematicheskaya Fizika}, 28:\penalty0
  341--352, 1976.
\newblock [English translation: Theoretical and Mathematical Physics, 1976,
  28:3, 829–837].

\bibitem[Kurkjian et~al.(2019)Kurkjian, Klimin, Tempere, and Castin]{higgs}
H.~Kurkjian, S.~N. Klimin, J.~Tempere, and Y.~Castin.
\newblock {Pair-Breaking Collective Branch in BCS Superconductors and
  Superfluid Fermi Gases}.
\newblock \emph{Phys. Rev. Lett.}, 122:\penalty0 093403, March 2019.
\newblock \doi{10.1103/PhysRevLett.122.093403}.
\newblock URL \url{https://link.aps.org/doi/10.1103/PhysRevLett.122.093403}.

\bibitem[Klimin et~al.(2019)Klimin, Tempere, and Kurkjian]{artlongsk}
S.~N. Klimin, J.~Tempere, and H.~Kurkjian.
\newblock {Phononic collective excitations in superfluid Fermi gases at nonzero
  temperatures}.
\newblock \emph{Phys. Rev. A}, 100:\penalty0 063634, December 2019.
\newblock \doi{10.1103/PhysRevA.100.063634}.
\newblock URL \url{https://link.aps.org/doi/10.1103/PhysRevA.100.063634}.

\bibitem[Patel et~al.(2019)Patel, Yan, Mukherjee, Fletcher, Struck, and
  Zwierlein]{Zwierlein2019}
Parth~B. Patel, Zhenjie Yan, Biswaroop Mukherjee, Richard~J. Fletcher, Julian
  Struck, and Martin~W. Zwierlein.
\newblock {Universal Sound Diffusion in a Strongly Interacting Fermi Gas}.
\newblock \textit{arXiv:1909.02555}, 2019.

\bibitem[Behrle et~al.(2018)Behrle, Harrison, Kombe, Gao, Link, Bernier,
  Kollath, and K{\"o}hl]{Koehl2018}
A.~Behrle, T.~Harrison, J.~Kombe, K.~Gao, M.~Link, J.~S. Bernier, C.~Kollath,
  and M.~K{\"o}hl.
\newblock {Higgs mode in a strongly interacting fermionic superfluid}.
\newblock \emph{Nature Physics}, 2018.
\newblock \doi{10.1038/s41567-018-0128-6}.
\newblock URL \url{https://doi.org/10.1038/s41567-018-0128-6}.

\bibitem[Pekker and Varma(2015)]{Varma2015}
David Pekker and C.M. Varma.
\newblock {Amplitude/Higgs Modes in Condensed Matter Physics}.
\newblock \emph{Annual Review of Condensed Matter Physics}, 6\penalty0
  (1):\penalty0 269--297, 2015.
\newblock \doi{10.1146/annurev-conmatphys-031214-014350}.
\newblock URL \url{https://doi.org/10.1146/annurev-conmatphys-031214-014350}.

\bibitem[Sooryakumar and Klein(1980)]{Klein1980}
R.~Sooryakumar and M.~V. Klein.
\newblock {Raman Scattering by Superconducting-Gap Excitations and Their
  Coupling to Charge-Density Waves}.
\newblock \emph{Phys. Rev. Lett.}, 45:\penalty0 660--662, August 1980.
\newblock \doi{10.1103/PhysRevLett.45.660}.
\newblock URL \url{https://link.aps.org/doi/10.1103/PhysRevLett.45.660}.

\bibitem[Matsunaga et~al.(2013)Matsunaga, Hamada, Makise, Uzawa, Terai, Wang,
  and Shimano]{Shimano2013}
Ryusuke Matsunaga, Yuki~I. Hamada, Kazumasa Makise, Yoshinori Uzawa, Hirotaka
  Terai, Zhen Wang, and Ryo Shimano.
\newblock {Higgs Amplitude Mode in the BCS Superconductors
  ${\mathrm{Nb}}_{1\mathrm{\text{\ensuremath{-}}}x}{\mathrm{Ti}}_{x}\mathbf{N}$
  Induced by Terahertz Pulse Excitation}.
\newblock \emph{Phys. Rev. Lett.}, 111:\penalty0 057002, July 2013.
\newblock \doi{10.1103/PhysRevLett.111.057002}.
\newblock URL \url{https://link.aps.org/doi/10.1103/PhysRevLett.111.057002}.

\bibitem[M\'easson et~al.(2014)M\'easson, Gallais, Cazayous, Clair, Rodi\`ere,
  Cario, and Sacuto]{Sacuto2014}
M.-A. M\'easson, Y.~Gallais, M.~Cazayous, B.~Clair, P.~Rodi\`ere, L.~Cario, and
  A.~Sacuto.
\newblock {Amplitude Higgs mode in the $2H\ensuremath{-}{\text{NbSe}}_{2}$
  superconductor}.
\newblock \emph{Phys. Rev. B}, 89:\penalty0 060503, February 2014.
\newblock \doi{10.1103/PhysRevB.89.060503}.
\newblock URL \url{https://link.aps.org/doi/10.1103/PhysRevB.89.060503}.

\bibitem[Cea et~al.(2015)Cea, Castellani, Seibold, and Benfatto]{Benfatto2015}
T.~Cea, C.~Castellani, G.~Seibold, and L.~Benfatto.
\newblock {Nonrelativistic Dynamics of the Amplitude (Higgs) Mode in
  Superconductors}.
\newblock \emph{Phys. Rev. Lett.}, 115:\penalty0 157002, October 2015.
\newblock \doi{10.1103/PhysRevLett.115.157002}.
\newblock URL \url{https://link.aps.org/doi/10.1103/PhysRevLett.115.157002}.

\bibitem[Grasset et~al.(2018)Grasset, Cea, Gallais, Cazayous, Sacuto, Cario,
  Benfatto, and M\'easson]{Measson2018}
Romain Grasset, Tommaso Cea, Yann Gallais, Maximilien Cazayous, Alain Sacuto,
  Laurent Cario, Lara Benfatto, and Marie-Aude M\'easson.
\newblock {Higgs-mode radiance and charge-density-wave order in
  $2H\ensuremath{-}{\mathrm{NbSe}}_{2}$}.
\newblock \emph{Phys. Rev. B}, 97:\penalty0 094502, March 2018.
\newblock \doi{10.1103/PhysRevB.97.094502}.
\newblock URL \url{https://link.aps.org/doi/10.1103/PhysRevB.97.094502}.

\bibitem[Grasset et~al.(2019)Grasset, Gallais, Sacuto, Cazayous, Ma\~nas
  Valero, Coronado, and M\'easson]{Measson2019}
Romain Grasset, Yann Gallais, Alain Sacuto, Maximilien Cazayous, Samuel Ma\~nas
  Valero, Eugenio Coronado, and Marie-Aude M\'easson.
\newblock Pressure-induced collapse of the charge density wave and higgs mode
  visibility in $2h\text{\ensuremath{-}}{\mathrm{tas}}_{2}$.
\newblock \emph{Phys. Rev. Lett.}, 122:\penalty0 127001, Mar 2019.
\newblock \doi{10.1103/PhysRevLett.122.127001}.
\newblock URL \url{https://link.aps.org/doi/10.1103/PhysRevLett.122.127001}.

\bibitem[Volovik and Zubkov(2014)]{Volovik2014}
G.~E. Volovik and M.~A. Zubkov.
\newblock {Higgs Bosons in Particle Physics and in Condensed Matter}.
\newblock \emph{Journal of Low Temperature Physics}, 175\penalty0 (1):\penalty0
  486--497, April 2014.
\newblock ISSN 1573-7357.
\newblock \doi{10.1007/s10909-013-0905-7}.
\newblock URL \url{https://doi.org/10.1007/s10909-013-0905-7}.

\bibitem[Abrosimov et~al.(2011)Abrosimov, Brink, Dellafiore, and
  Matera]{Matera2010}
V.I. Abrosimov, D.M. Brink, A.~Dellafiore, and F.~Matera.
\newblock Self-consistency and search for collective effects in semiclassical
  pairing theory.
\newblock \emph{Nuclear Physics A}, 864\penalty0 (1):\penalty0 38 -- 62, 2011.
\newblock ISSN 0375-9474.
\newblock \doi{https://doi.org/10.1016/j.nuclphysa.2011.06.020}.
\newblock URL
  \url{http://www.sciencedirect.com/science/article/pii/S0375947411004441}.

\bibitem[Abrosimov et~al.(2014)Abrosimov, Brink, and Matera]{Matera2014}
V.~I. Abrosimov, D.~M. Brink, and F.~Matera.
\newblock Pairing collective modes in superfluid nuclei: a semiclassical
  approach.
\newblock \emph{Bulletin of the Russian Academy of Sciences: Physics},
  78\penalty0 (7):\penalty0 630--633, 2014.

\bibitem[Castin and Kurkjian(2019)]{higgslong}
Y.~Castin and H~Kurkjian.
\newblock {Collective excitation branch in the continuum of pair-condensed
  Fermi gases: analytical study and scaling laws}.
\newblock \textit{arXiv:1907.12238}, 2019.

\bibitem[Tsuchiya et~al.(2013)Tsuchiya, Ganesh, and Nikuni]{Nikuni2013}
Shunji Tsuchiya, R.~Ganesh, and Tetsuro Nikuni.
\newblock {Higgs mode in a superfluid of Dirac fermions}.
\newblock \emph{Phys. Rev. B}, 88:\penalty0 014527, July 2013.
\newblock \doi{10.1103/PhysRevB.88.014527}.
\newblock URL \url{https://link.aps.org/doi/10.1103/PhysRevB.88.014527}.

\bibitem[Bruun(2014)]{Bruun2014}
G.~M. Bruun.
\newblock {Long-lived Higgs mode in a two-dimensional confined Fermi system}.
\newblock \emph{Phys. Rev. A}, 90:\penalty0 023621, August 2014.
\newblock \doi{10.1103/PhysRevA.90.023621}.
\newblock URL \url{https://link.aps.org/doi/10.1103/PhysRevA.90.023621}.

\bibitem[Anderson(1958)]{Anderson1958}
P.W. Anderson.
\newblock {Random-Phase Approximation in the Theory of Superconductivity}.
\newblock \emph{Phys. Rev.}, 112:\penalty0 1900--1916, 1958.

\bibitem[Kurkjian and Tempere(2017)]{artmicro}
Hadrien Kurkjian and Jacques Tempere.
\newblock {Absorption and emission of a collective excitation by a fermionic
  quasiparticle in a Fermi superfluid}.
\newblock \emph{New Journal of Physics}, 19\penalty0 (11):\penalty0 113045,
  2017.
\newblock URL \url{http://stacks.iop.org/1367-2630/19/i=11/a=113045}.

\bibitem[Beliaev(1958)]{Beliaev1958}
S.T. Beliaev.
\newblock {Application of the Methods of Quantum Field Theory to a System of
  Bosons}.
\newblock \emph{Zh. Eksp. Teor. Fiz.}, 34:\penalty0 417, August 1958.

\bibitem[Landau and Khalatnikov(1949)]{Khalatnikov1949}
Lev Landau and Isaak Khalatnikov.
\newblock {Teoriya vyazkosti {Geliya-II}}.
\newblock \emph{Zh. Eksp. Teor. Fiz.}, 19:\penalty0 637, 1949.

\bibitem[Kurkjian et~al.(2017)Kurkjian, Castin, and Sinatra]{annalen}
H.~Kurkjian, Y.~Castin, and A.~Sinatra.
\newblock {Three-Phonon and Four-Phonon Interaction Processes in a
  Pair-Condensed Fermi Gas}.
\newblock \emph{Annalen der Physik}, 529\penalty0 (9):\penalty0 1600352, 2017.
\newblock ISSN 1521-3889.
\newblock \doi{10.1002/andp.201600352}.
\newblock URL \url{http://dx.doi.org/10.1002/andp.201600352}.

\bibitem[Wong and Takada(1988)]{Takada1988}
K.~Y.~M. Wong and S.~Takada.
\newblock Effects of quasiparticle screening on collective modes. ii.
  superconductors.
\newblock \emph{Phys. Rev. B}, 37:\penalty0 5644--5656, Apr 1988.
\newblock \doi{10.1103/PhysRevB.37.5644}.
\newblock URL \url{https://link.aps.org/doi/10.1103/PhysRevB.37.5644}.

\bibitem[Bruun and Mottelson(2001)]{Mottelson2001}
G.~M. Bruun and B.~R. Mottelson.
\newblock {Low Energy Collective Modes of a Superfluid Trapped Atomic Fermi
  Gas}.
\newblock \emph{Phys. Rev. Lett.}, 87:\penalty0 270403, December 2001.
\newblock \doi{10.1103/PhysRevLett.87.270403}.
\newblock URL \url{https://link.aps.org/doi/10.1103/PhysRevLett.87.270403}.

\bibitem[Minguzzi et~al.(2001)Minguzzi, Ferrari, and Castin]{Castin2001}
A.~Minguzzi, G.~Ferrari, and Y.~Castin.
\newblock {Dynamic structure factor of a superfluid Fermi gas}.
\newblock \emph{The European Physical Journal D - Atomic, Molecular, Optical
  and Plasma Physics}, 17\penalty0 (1):\penalty0 49--55, October 2001.
\newblock ISSN 1434-6079.
\newblock \doi{10.1007/s100530170036}.
\newblock URL \url{https://doi.org/10.1007/s100530170036}.

\bibitem[He(2016)]{He2016}
Lianyi He.
\newblock {Dynamic density and spin responses of a superfluid Fermi gas in the
  BCS–BEC crossover: Path integral formulation and pair fluctuation theory}.
\newblock \emph{Annals of Physics}, 373:\penalty0 470 -- 511, 2016.
\newblock ISSN 0003-4916.
\newblock \doi{https://doi.org/10.1016/j.aop.2016.07.030}.
\newblock URL
  \url{http://www.sciencedirect.com/science/article/pii/S0003491616301312}.

\bibitem[Marini et~al.(1998)Marini, Pistolesi, and Strinati]{Strinati1998}
M.~Marini, F.~Pistolesi, and G.C. Strinati.
\newblock {Evolution from BCS superconductivity to Bose condensation: analytic
  results for the crossover in three dimensions}.
\newblock \emph{European Physical Journal B}, 1:\penalty0 151--159, 1998.

\bibitem[Combescot et~al.(2006)Combescot, Kagan, and Stringari]{CKS2006}
R.~Combescot, M.~Yu. Kagan, and S.~Stringari.
\newblock {Collective mode of homogeneous superfluid Fermi gases in the BEC-BCS
  crossover}.
\newblock \emph{Phys. Rev. A}, 74:\penalty0 042717, October 2006.
\newblock \doi{10.1103/PhysRevA.74.042717}.
\newblock URL \url{http://link.aps.org/doi/10.1103/PhysRevA.74.042717}.

\bibitem[Kulik et~al.(1981)Kulik, Entin-Wohlman, and Orbach]{Orbach1981}
I.~O. Kulik, Ora Entin-Wohlman, and R.~Orbach.
\newblock {Pair susceptibility and mode propagation in superconductors: A
  microscopic approach}.
\newblock \emph{Journal of Low Temperature Physics}, 43\penalty0 (5):\penalty0
  591--620, June 1981.
\newblock ISSN 1573-7357.
\newblock \doi{10.1007/BF00115617}.
\newblock URL \url{https://doi.org/10.1007/BF00115617}.

\bibitem[Ohashi and Takada(1997)]{Takada1997}
Yoji Ohashi and Satoshi Takada.
\newblock {Goldstone Mode in Charged Superconductivity: Theoretical Studies of
  the Carlson-Goldman Mode and Effects of the Landau Damping in the
  Superconducting State}.
\newblock \emph{Journal of the Physical Society of Japan}, 66\penalty0
  (8):\penalty0 2437--2458, 1997.
\newblock \doi{10.1143/JPSJ.66.2437}.

\bibitem[Gurarie(2009)]{Gurarie2009}
V.~Gurarie.
\newblock {Nonequilibrium Dynamics of Weakly and Strongly Paired
  Superconductors}.
\newblock \emph{Phys. Rev. Lett.}, 103:\penalty0 075301, August 2009.
\newblock \doi{10.1103/PhysRevLett.103.075301}.
\newblock URL \url{https://link.aps.org/doi/10.1103/PhysRevLett.103.075301}.

\bibitem[Kurkjian(2016)]{TheseHK}
H.~Kurkjian.
\newblock \emph{{Cohérence}, brouillage et dynamique de phase dans un
  condensat de paires de fermions}.
\newblock PhD thesis, \'Ecole Normale Supérieure, Paris, 2016.

\bibitem[Ohashi and Griffin(2003)]{Griffin2003}
Y.~Ohashi and A.~Griffin.
\newblock {Superfluidity and collective modes in a uniform gas of Fermi atoms
  with a Feshbach resonance}.
\newblock \emph{Phys. Rev. A}, 67:\penalty0 063612, June 2003.
\newblock \doi{10.1103/PhysRevA.67.063612}.
\newblock URL \url{https://link.aps.org/doi/10.1103/PhysRevA.67.063612}.

\end{thebibliography}
\bibliographystyle{unsrtnat}

\end{document}